\documentclass[prl,twocolumn,superscriptaddress]{revtex4-2}
\usepackage{url}
\pdfoutput=1 

\usepackage[T1]{fontenc} 

\usepackage{epsfig}
\usepackage{amsfonts}
\usepackage{graphicx}
\usepackage{epsfig}
\usepackage{eepic}
\usepackage{amsmath}
\usepackage{amssymb}
\usepackage{color}
\usepackage{bbm}
\usepackage{dcolumn}
\usepackage{bm}
\usepackage{ulem}
\usepackage{mathrsfs}
\usepackage{bbold}
\usepackage{datetime}

\usepackage{overpic} 
\usepackage{rotating}
\usepackage[usenames,dvipsnames]{xcolor}
\usepackage[colorlinks=true,citecolor=Blue,linkcolor=Green,urlcolor=Green]{hyperref}

\def\bc{\begin{center}}

\def\ec{\end{center}}
\def\be{\begin{eqnarray}}
\def\ee{\end{eqnarray}}
\definecolor{dyellow}{rgb}{1.,0.8,.0}
\definecolor{myblue}{rgb}{.1,.1,.7}
\definecolor{dcyan}{rgb}{.0,.6,.6}
\definecolor{dmagenta}{rgb}{0.6,0.0,0.6}
\definecolor{brown}{rgb}{0.6,0.2,0.}
\definecolor{darkblue}{rgb}{.0,.0,0.5}
\definecolor{darkred}{rgb}{0.75,0.0,0.0}
\definecolor{orange}{rgb}{1.,.6,.0}
\definecolor{dorange}{rgb}{0.8,.4,.0}
\definecolor{darkgreen}{rgb}{0.0,0.6,0.0}
\definecolor{purple}{rgb}{.4,.0,.4}
\definecolor{lightgrey}{rgb}{0.7, 0.7, 0.7}
\definecolor{grey}{rgb}{0.4, 0.4, 0.4}

\newcommand{\xdownarrow}[1]{%
  {\left\downarrow\vbox to #1{}\right.\kern-\nulldelimiterspace}
}
\newcommand{\xuparrow}[1]{%
  {\left\uparrow\vbox to #1{}\right.\kern-\nulldelimiterspace}
}

\begin{document}

\title{Geometric-phase control of Krylov complexity in adiabatic dynamics}

\author{Han-Qi Zheng}
\email{by2530012@buaa.edu.cn}
\affiliation{Center for Gravitational Physics, Department of Space Science, Beihang University, Beijing 100191, China}
\author{Peng-Zhang He}\email{hepzh@cwnu.edu.cn}
\affiliation{School of Physics and Astronomy, China West Normal University, Nanchong 637002, Sichuan, China}
\author{Lei-Hua Liu}\email{liuleihua8899@hotmail.com}
\affiliation{Department of Physics, College of Physics, Mechanical and Electrical Engineering, Jishou University, Jishou 416000, China}
\author{Hai-Qing Zhang}
\email{hqzhang@buaa.edu.cn}
\affiliation{Center for Gravitational Physics, Department of Space Science, Beihang University, Beijing 100191, China}
\affiliation{Peng Huanwu Collaborative Center for Research and Education, Beihang University, Beijing
100191, China}

\begin{abstract}
We show that geometric phases accumulated during adiabatic evolution can be converted into observable interference in Krylov space, leading to a geometric-phase-dependent Krylov oscillation. Adiabatic dynamics force Krylov complexity to vanish if the initial Krylov basis is an instantaneous eigenstate of the Hamiltonian; Nevertheless, we demonstrate that the Krylov complexity will be non-vanishing if the initial Krylov basis is a superposition state rather than an eigenstate. Consequently, Krylov complexity is found to depend on the difference of dynamical phases in the Krylov space, which is deeply related to the Berry connections in the original Hilbert space. In particular, for a single qubit system with constant Lanczos coefficients, Krylov complexity oscillates harmonically at a frequency given by the strength of the external field and the geometric Berry phase. Therefore, our work may provide a novel avenue to probe the geometric phase from the Krylov complexity. 
\end{abstract}

\maketitle

\paragraph{Introduction ---}
Quantum complexity has established a bridge between quantum information, many-body physics, quantum field theory and quantum gravity, providing an important diagnostic method for tracking the dynamics of quantum operators and quantum states during time evolution \cite{Chapman:2021jbh,Baiguera:2025dkc}. Krylov complexity, as one of the indicators for measuring quantum complexity, has seen rapid development in recent years because of its significance in the modern physics \cite{Nandy:2024evd,Rabinovici:2025otw}. 

Krylov complexity was initially proposed to quantify the operator spreading and diagnose the quantum chaos in many-body systems \cite{Parker:2018yvk}. Later, it was extended to the state-oriented spread complexity, adapting the formalism to Schr\"odinger state evolution \cite{Balasubramanian:2022tpr}. Subsequent research has extended it to many branches of theoretical physics: In the holographic framework, large number of investigations have characterized the operator growth in the Double-Scaled SYK model and JT gravity, uncovering the universal scalings in strongly coupled chaotic systems \cite{Rabinovici:2023yex,Ambrosini:2024sre,Lin:2022rbf}; Studies in conformal field theories further established the bound on chaos and an exponential growth of Krylov complexity \cite{Dymarsky:2021bjq,Kundu:2023hbk}; In many-body physics, Krylov observables have efficiently discriminated between the ergodic systems and many-body localizations \cite{Dymarsky:2019elm,Trigueros:2021rwj,Rabinovici:2021qqt, Rabinovici:2022beu}. Aside from these notable breakthroughs, Krylov complexity has also developed significantly in other directions: the connections between the coherent states and symmetries in Krylov subspace \cite{Caputa:2021sib,Zhai:2024tkz,Grabarits:2026hjz}, Krylov complexity in open quantum systems \cite{Bhattacharya:2022gbz,Liu:2022god,Bhattacharya:2023zqt}, in quantum field theories \cite{Adhikari:2022whf,Avdoshkin:2022xuw,Camargo:2022rnt, Vasli:2023syq,He:2024xjp,He:2024hkw,He:2025guu}, in qubit systems \cite{Aguilar-Gutierrez:2023nyk,Caputa:2024vrn,Seetharaman:2024ket}, in early universe \cite{Li:2024kfm, Li:2024iji, Zhai:2024odw}, and the relation to other types of complexities, such as Nielsen complexity and circuit complexity \cite{Aguilar-Gutierrez:2023nyk,Craps:2023ivc,Craps:2025kub,Lv:2023jbv}. There are indeed a bunch of important works on Krylov complexity that we cannot all list. Interested readers may refer to the seminal review papers \cite{Nandy:2024evd,Rabinovici:2025otw}.

Krylov complexity was previously introduced with time-independent Hamiltonians \cite{Parker:2018yvk}, and was constructed based on the Lanczos tridiagonalization on the Krylov subspaces generated by the repeated actions of the Hamiltonian operator \cite{viswanath1994recursion}. However, extending it to time-dependent Hamiltonian is challenging since Hamiltonian operator $\hat H(t)$ with different time is usually non-commutative. Therefore, the previous algorithm for Krylov subspace does not apply. Thanks to \cite{Takahashi:2024hex}, the authors introduced the $(t, t')$ formalism to successfully build the Krylov algorithm with time-dependent Hamiltonian \cite{Howland:1974aa,Peskin:1993,Peskin:1994}. In this case, Krylov subspace is generated by the operator $(\hat H(t)-i\partial_t)$, such that $b_{k+1}(t)|K_{k+1}(t)\rangle=(\hat H(t)-i\partial_t)|K_k(t)\rangle-a_k(t)|K_k(t)\rangle-b_k(t)|K_{k-1}(t)\rangle$ and the Krylov basis $\{|K_k(t)\rangle\}_{k=0}^{d-1}$ satisfying the orthonormal condition $\langle K_m(t)|K_n(t)\rangle=\delta_{mn}$ with $d$ the dimension of the Krylov subspace. The Lanczos coefficients $\{a_k(t)\}_{k=0}^{d-1}$ and  $\{b_k(t)\}_{k=1}^{d-1}$ are real, and form the matrix $\mathcal{\hat L}=\sum_{k=0}^{d-1}a_k(t)|k\rangle\langle k|+\sum_{k=1}^{d-1}b_k(t)(|k\rangle\langle{k\!-\!1}|+|{k\!-\!1}\rangle\langle k|)$. Then the state Krylov complexity (spread complexity) can be defined as 
\be\label{defK} K(t)=\sum_{k=0}^{d-1}k\lvert\varphi_k(t)\rvert^2, \ee 
where $\varphi_k(t)=\langle k|\varphi(t)\rangle$. The transformed state $|\varphi(t)\rangle$ satisfies the equation $i\partial_t|\varphi(t)\rangle=\mathcal{\hat L}(t)|\varphi(t)\rangle$ with the initial condition $|\varphi(0)\rangle=|0\rangle$. The above definition of Krylov complexity is equivalent to the expectation value of the Krylov operator $\mathcal{\hat K}$, such that $K(t)=\langle\varphi(t)|\mathcal{\hat K}|\varphi(t)\rangle$ with $\mathcal{\hat K}=\sum_{k=0}^{d-1}k|k\rangle\langle k|$. Other schemes for time-dependent Hamiltonian can be found in \cite{Grabarits:2026hjz,FarajiAstaneh:2026aks}.

Although Krylov complexity with time-dependent Hamiltonian has been constructed, the Krylov complexity in adiabatic dynamics is still underdeveloped. At first sight, it seems that Krylov complexity will always be zero, since the state will remain in the instantaneous eigenstate in adiabatic dynamics. This is true if the initial Krylov basis is exactly the instantaneous eigenstate of the Hamiltonian. However, if one chooses the initial Krylov basis as a superposition state rather than an eigenstate of the Hamiltonian, Krylov complexity will not vanish! This is the key point in this letter. Berry phase as a geometric phase is a vital concept in quantum adiabatic dynamics \cite{Berry:1984jv,shapere1989geometric}. It has wide applications in quantum Hall effect \cite{Thouless:1982zz,KOHMOTO1985343}, anomalous Hall effect \cite{Jungwirth:2002zz}, topological insulators \cite{RevModPhys.82.3045} and so on \cite{Xiao:2009rm}. Therefore, to explore the connections between the Krylov complexity and the geometric phase in adiabatic dynamics is an urgent and tantalizing task.

\paragraph{Krylov subspace in adiabatic dynamics ---}
We consider a single spin system with a time-dependent Hamiltonian
\be\label{hamiltonian} \hat H(t)=h(t)\hat{\bf S}\cdot{\bf n}(t) ,\ee 
where $\hat{\bf S}=(\hat S_x, \hat S_y, \hat S_z)$ is the spin operator, ${\bf n}(t)=(n_x, n_y, n_z)=\big(\sin\theta(t)\cos\varphi(t),\sin\theta(t)\sin\varphi(t),\cos\theta(t)\big)$ is the unit vector and $h(t)$ is the strength of the external field such as magnetic field. We will focus on the adiabatic dynamics of the system. Therefore, we adopt the instantaneous eigenstate $|m(t)\rangle$ as basis such that $\hat H(t)|m(t)\rangle=h(t)m|m(t)\rangle$ in which $m$ is the eigenvalue satisfying $m=-S, -S+1,\cdots, S$ with $S$ the maximum spin.  In adiabatic dynamics, the instantaneous eigenstate obeys the following equation of motion \cite{Sakurai:2011zz}, 
\be\label{evolve}
i\partial_t|m(t)\rangle=|m(t)\rangle i\langle m(t)|\dot m(t)\rangle,
\ee
in which the symbol $\dot{}$ indicates the time derivative. Here, $i\langle m(t)|\dot m(t)\rangle\!=\!i\langle m({\bf R}(t))|\nabla_{\bf R} m({\bf R}(t))\rangle\cdot\!{\bf \dot R}(t)$ where $i\langle m({\bf R}(t))|\nabla_{\bf R} m({\bf R}(t))\rangle$ is the {\it Berry connection} or {\it Berry vector potential}, and ${\bf\dot R}(t)$ is the velocity in the parameters' space. From the Krylov space generator $\hat H(t)-i\partial_t$, it is natural to expect that the Krylov complexity in adiabatic evolutions will be relevant to the Berry connection, and subsequently Berry phase if the system evolves in a closed path. 
Further computations of Eq.\eqref{evolve} lead to
$
i\partial_t|m(t)\rangle=-m\dot\varphi(1-\cos\theta)|m(t)\rangle\equiv A_m(t)|m(t)\rangle 
$ \cite{SM}. Therefore, the information of Berry connection has been encoded in $A_m$.

In existing literatures, people mainly build the Krylov subspace out of the eigenstates of the Hamiltonian \cite{Takahashi:2024hex}. In adiabatic systems, if we set the initial Krylov basis as the instantaneous eigenstate, such as $|K_0(t)\rangle=|m(t)\rangle$, we will get $a_0=\langle K_0\lvert(\hat H-i\partial_t)\rvert K_0\rangle=m(h+\dot\varphi(1-\cos\theta))$ and $b_1|K_1\rangle=(\hat H-i\partial_t)|K_0\rangle-a_0|K_0\rangle=0$. From the theory of Krylov subspace, higher order Krylov bases are also vanishing. Therefore, one cannot construct non-trivial Krylov subspace in this sense and the Krylov complexity is obviously zero. The above computation can be understood intuitively under this assumption: the initial Krylov basis we choose will always stay in the $m$-th eigenstate as time evolves, thus there will be no transitions between energy levels, which leads to vanishing Krylov complexity.

However, if we choose the initial Krylov basis as a superposition state rather than an eigenstate, situations will be different. Without loss of generality, we set the initial Krylov basis as a superposition state of two eigenstates of a single spin system,
\be \label{initial}
|K_0(t)\rangle=\cos\left(\frac{\alpha(t)}{2}\right)|m(t)\rangle+\sin\left(\frac{\alpha(t)}{2}\right)e^{i\beta(t)}|n(t)\rangle,~
\ee 
where $m\neq n$, and $\alpha(t), \beta(t)$ are real functions in time.  From the algorithm developed in \cite{Takahashi:2024hex}, the Krylov basis $\{|K_k(t)\rangle\}$ and the Lanczos coefficients $\{a_k\}, \{b_k\}$ can be readily obtained: $b_0=0$, $a_0(t)=\left(mh-A_m\right)\cos^2\left(\frac{\alpha}{2}\right)+(nh-A_n+\dot\beta)\sin^2\left(\frac{\alpha}{2}\right)$ and 
\be
\begin{cases}
b_1(t)=\frac12\sqrt{\xi^2\sin^2\alpha+\dot\alpha^2}, \\
a_1(t)=\frac{1}{8b_1^2}\big[(\dot\alpha^2\!+\!\xi^2 \sin^2\alpha)\big((m+n)h\!-\!A_m\!-\!A_n\!+\!\dot\beta \\ 
 \hspace{1.cm} -\xi\cos \alpha\big)\!-\!2\dot\alpha \dot \xi \sin \alpha 
 \!+\!2\xi (\ddot\alpha \sin  \alpha \!-\!\dot\alpha^2 \cos\alpha)\big],\\
|K_1(t)\rangle=\frac{\xi\sin\alpha+i\dot\alpha}{2b_1}\big[\sin(\frac{\alpha}{2})|m(t)\rangle 
 \!-\!\cos(\frac{\alpha}{2})e^{i\beta}|n(t)\rangle\big],
\end{cases}\!\!\!
\ee
in which $\xi(t)\equiv h(m-n)-A_m+A_n-\dot\beta$.
It is easy to check that $\langle K_0(t)|K_1(t)\rangle=0$. Higher orders of the Krylov bases will vanish since $b_2|K_2(t)\rangle=(\hat H-i\partial_t)|K_1\rangle-a_1|K_1\rangle-b_1|K_0\rangle=0$.  
Therefore, we have constructed a two-dimensional Krylov subspace spanned by $\{|K_0(t)\rangle, |K_1(t)\rangle\}$ in adiabatic dynamics and the Lanczos matrix is 
\be
\mathcal{\hat L}\!=\!\!\sum_{k=0}^{1}a_k(t)|k\rangle\langle k|\!+\!b_1(t)(|1\rangle\langle {0}|\!+\!|{0}\rangle\langle 1|)
\!=\!\begin{pmatrix}
a_0(t)\!&b_1(t)\\
b_1(t)\!&a_1(t)
\end{pmatrix}.
\ee

\begin{figure}[t]
\centering
\includegraphics[trim=0.cm 0.cm 0cm 0cm, clip=true, scale=0.35]{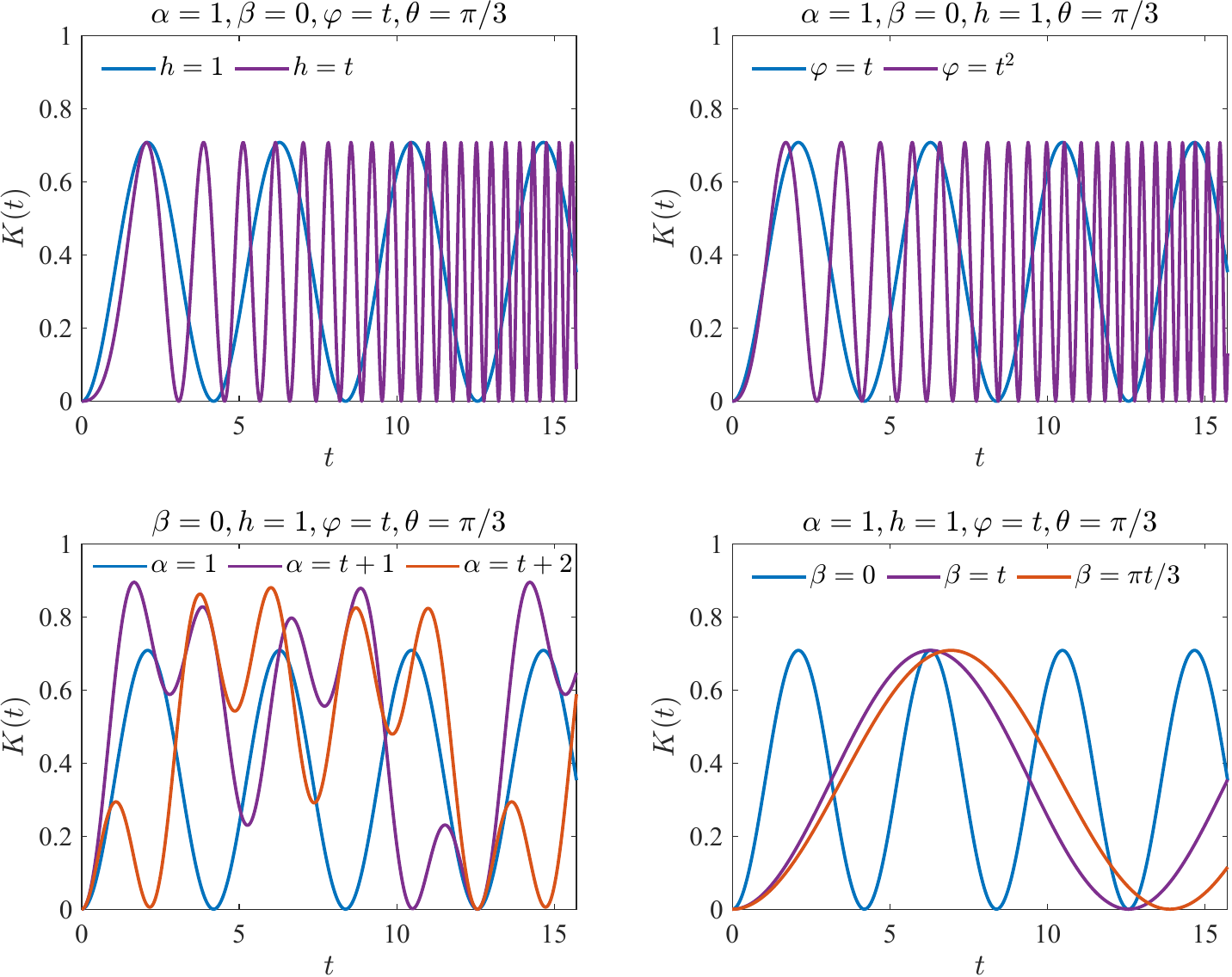}
\caption{Numerical results of the time evolution of Krylov complexity in a single qubit system with various time-dependent parameters. The blue lines are the reference lines with constant controlling parameters.}\label{p1}
\end{figure}

\paragraph{Krylov complexity for a single qubit system ---}
 For simplicity we will focus on the adiabatic dynamics of a $S=1/2$ spin system -- a two-level single qubit system. In Fig.\ref{p1} we show the evolution of the Krylov complexity with various controlling parameters. In these four panels, the blue lines oscillate harmonically in time, indicating that the system is an autonomous system with constant parameters \footnote{For the azimuthal angle $\varphi(t)$, we always set it time-dependent.  What we mean the constant parameter for $\varphi(t)$ is that its angular velocity $\omega=\dot\varphi(t)$ is a constant rather than $\varphi(t)$ itself.}. In the upper two panels, $K(t)$ has similar behaviors for $h(t)=t$ and $\varphi(t)=t^2$, such that as time evolves, the lines in purple become more and more dense. It indicates that the external field strength $h$ and the angular velocity $\omega\equiv\dot\varphi$ may have similar impacts on the frequency of the Krylov complexity.  
 This will become apparent after we derive the analytical results of the Krylov complexity later. 
 In the lower left panel of Fig.\ref{p1}, Krylov complexity becomes quasi-periodic if $\alpha(t)$ is a linear function on time, since the system now is no longer autonomous \cite{Aguilar-Gutierrez:2023nyk,Caputa:2024vrn,Seetharaman:2024ket}. On the contrary, from the lower right panel, the linear dependence on time for $\beta(t)$ seems not change the periodic behavior of the Krylov complexity.  
 
\paragraph{A closed formalism of Krylov complexity ---} 
From the equation $i\partial_t|\varphi(t)\rangle=\mathcal{\hat L}|\varphi(t)\rangle$ and setting $|\varphi(t)\rangle=(c_0(t),c_1(t))^\text{T}$ (where $^\text{T}$ indicates transposing of the vector),  we get 
\be\label{scheq}
i\begin{pmatrix}
\dot c_0(t)\\
\dot c_1(t)
\end{pmatrix}=\begin{pmatrix}
a_0(t)&b_1(t)\\
b_1(t)&a_1(t)
\end{pmatrix}\begin{pmatrix}
c_0(t)\\
c_1(t)
\end{pmatrix}.
\ee
In the following, we will denote $\mathcal{\hat L}$ as the effective Hamiltonian $H_\text{eff}$ in the Krylov subspace, i.e., $H_{\text{eff}}\equiv\mathcal{\hat L}$. 
In principle, one can diagonalize $H_{\text{eff}}$ to solve the equation \eqref{scheq}. But at first, we can analytically solve it if the coefficients are time-independent, i.e., $a_0(t)=a_0, a_1(t)=a_1, b_1(t)=b_1$ are constants. Then, we can readily get \cite{SM}
\be\label{Ktconst}
K(t)&=&|c_1(t)|^2=\frac{4b_1^2\sin ^2\left(\frac{1}{2} t \sqrt{(a_0-a_1)^2+4b_1^2}\right)}{(a_0-a_1)^2+4b_1^2}\nonumber\\
&=&\sin^2\alpha\,\sin^2\left(\frac{h+\dot\varphi(1-\cos\theta)}{2}t\right).~~~
\ee
which resembles the {\it Rabi oscillations} in two-level systems \cite{Rabi:1937dgo}.
Given that $2\sin^2(x/2)=1-\cos x$, the frequency of the oscillation is 
$
 \nu=h+\dot\varphi(1-\cos\theta).
$
 Therefore, it becomes clearer now why $h$ and $\dot\varphi$ have similar effects on the frequency in Fig.\ref{p1} as we discussed before, although we derive it from the constant parameters here. 
Moreover, we can consider the external field to be rotated slowly along the $z$-axis in a closed path. Since in this case the angular velocity $\omega=\dot\varphi$ is a constant, the periodicity $T$ equals $T=2\pi/\omega$. Thus, the second term in the frequency $\nu$ becomes $\dot\varphi(1-\cos\theta)=2\pi(1-\cos\theta)/T=\Omega_{\text{solid}}/T$, in which $\Omega_{\text{solid}}$ is the solid angle
$ \Omega_{\text{solid}}=\int_0^{2\pi} d\varphi\int_0^\theta d\theta'\sin\theta'=2\pi(1-\cos\theta)$.
This solid angle is exactly related to the Berry phase as $\gamma_{B}=-\frac12\Omega_{\text{solid}}$ \cite{Sakurai:2011zz}. Therefore, the frequency of the oscillation of the Krylov complexity becomes
\be \label{frequency2}
\nu=h+\Omega_{\text{solid}}/T=h-2\gamma_{B}/T. 
\ee 
It is seen that the frequency depends on the strength of the external field $h$ and the Berry phase $\gamma_{B}$ as well as the periodicity $T$. The oscillation frequency determined by the external field strength $h$ originates from the conventional dynamical phase accumulation, analogous to the {\it Larmor precession} of a spin in a static magnetic field \cite{Sakurai:2011zz}. By contrast, the Berry phase contribution $-2\gamma_B/T$ encodes the cumulative geometric effect of the adiabatic cyclic parameter evolution. Therefore, as the system adiabatically follows the instantaneous eigenstates while tracing a closed path, the accumulated Berry phase acts as an effective geometric ``driving term'' that shifts the oscillation frequency of the Krylov complexity.

However, if the parameters are time-dependent, it is difficult to solve the equation \eqref{scheq}. We will first diagonalize the effective Hamiltonian $H_{\text{eff}}$. The eigenvalues of $H_{\text{eff}}$ is 
$
 \lambda_\pm(t)=\frac{a_0(t)+a_1(t)}{2}\pm\frac12\Omega,
$
in which $\Omega\equiv\sqrt{\Delta(t)^2+4b_1(t)^2}$ and $\Delta(t)\equiv a_0(t)-a_1(t)$. Then, the effective Hamiltonian $H_{\text{eff}}$ can be transformed to the diagonal matrix $\Lambda(t)$ as
$
H_{\text{eff}}=U(t)\Lambda(t)U^\dagger(t)
$, 
where $\Lambda(t)=\text{diag}(\lambda_+(t), \lambda_-(t))$
and $U(t)$ is the transforming matrix $U(t)=\begin{pmatrix}
|{+}(t)\rangle,|{-}(t)\rangle
\end{pmatrix}$,
in which $|{\pm}(t)\rangle$  are the instantaneous eigenstates of the effective Hamiltonian,
$ H_{\text{eff}}|{\pm}(t)\rangle=\lambda_\pm(t)|{\pm}(t)\rangle.$
The state $|\varphi(t)\rangle$ can be expanded in the basis $|0\rangle= {1\choose0}, |1\rangle={0\choose1}$ as 
\be
|\varphi(t)\rangle=\begin{pmatrix}
c_0(t)\\
c_1(t)
\end{pmatrix}=c_0(t)|0\rangle+c_1(t)|1\rangle.
\ee
Therefore, the Krylov complexity becomes $K(t)=\langle\varphi(t)|\mathcal{\hat K}|\varphi(t)\rangle=|c_1(t)|^2$. On the other side, we can also expand $|\varphi(t)\rangle$ in the instantaneous eigenstates $|{\pm}(t)\rangle$ as
\be\label{15}
|\varphi(t)\rangle=d_+(t)|{+}(t)\rangle+d_-(t)|{-}(t)\rangle.~~
\ee
Let's assume ${\bf c}(t)=(c_0(t), c_1(t))^\text{T}$, therefore, it connects to the vector ${\bf d}(t)=(d_+(t),d_-(t))^\text{T}$ as ${\bf c}(t)=U(t){\bf d}(t)$. Hence, from the equation  $i\partial_t|\varphi(t)\rangle=\mathcal{\hat L}|\varphi(t)\rangle$, we get
$
i\dot{\bf c}(t)=i\dot U(t){\bf d}(t)+iU(t)\dot{\bf d}(t)=H_{\text{eff}}U(t){\bf d}(t)=U(t)\Lambda(t)U^\dagger(t)U(t){\bf d}(t)=U(t)\Lambda(t){\bf d}(t).
$
Multiply $U^\dagger(t)$ from the left, we obtain
\be\label{sch2}
i\dot{\bf d}(t)=\Lambda(t){\bf d}(t)-i{\bf A}(t){\bf d}(t),
\ee
where ${\bf A}(t)\equiv iU^\dagger(t)\dot U(t)$ which can be explicitly written as
\be
{\bf A}(t)=i\begin{pmatrix}
\langle+|\dot+\rangle&\langle+|\dot-\rangle\\
\langle-|\dot+\rangle&\langle-|\dot-\rangle
\end{pmatrix}.
\ee
The diagonal terms are the {\it Berry connections} from the instantaneous eigenstates of $H_\text{eff}$, while the non-diagonal terms indicate the non-adiabatic couplings. Therefore, ${\bf A}(t)$ is related to the  Berry connections in the Krylov space \footnote{We should stress that ${\bf A}(t)$ is not the Berry connection matrix from the instantaneous eigenstates of original Hamiltonian \eqref{hamiltonian}, but rather it is from the instantaneous eigenstates of the effective Hamiltonian $H_{\text{eff}}$ composed of the Lanczos coefficients.}! 

To analytically solve Eq.\eqref{sch2} is formidable. We further assume the adiabatic approximations in the Krylov space as well, such that $|\langle+|\dot-\rangle|\ll|\lambda_+{-}\lambda_-|$ and $|\langle-|\dot+\rangle|\ll|\lambda_+{-}\lambda_-|$. Therefore, we can neglect the non-diagonal terms in ${\bf A}(t)$,
\be
{\bf A}(t)\approx i\begin{pmatrix}
\langle+|\dot+\rangle&0\\
0&\langle-|\dot-\rangle
\end{pmatrix}.
\ee
Substitute it to the Eq.\eqref{sch2}, we get 
$i\dot d_\pm(t)\approx(\lambda_\pm(t)-i\langle\pm|\dot\pm\rangle)d_\pm(t)$.  
The integrals are easily obtained \cite{Sakurai:2011zz}
\be
d_\pm(t)&=& d_\pm(0)e^{i\theta_\pm(t)}e^{i\gamma_\pm(t)},
\ee
in which $\theta_\pm(t)\equiv-\int_0^t\lambda_\pm(\tau)d\tau$ is the dynamical phase in the Krylov space, while $\gamma_\pm(t)\equiv i\int_0^t\langle\pm(\tau)|\dot \pm(\tau)\rangle d\tau$ is the Berry phase in the Krylov space. 

At the initial time, $c_0(0)=1, c_1(0)=0$, thus, 
$
|\varphi(0)\rangle=|0\rangle=d_+(0)|{+}(0)\rangle+d_-(0)|{-}(0)\rangle.
$
Therefore, 
\be\label{d+d-}
d_\pm(0)=\langle{\pm}(0)|\varphi(0)\rangle=\langle{\pm}(0)|0\rangle.
\ee
Hence,  from Eq.\eqref{15} the solution of $|\varphi(t)\rangle$ is
$
|\varphi(t)\rangle=d_+(0)e^{i\theta_+(t)}e^{i\gamma_+(t)}|{+}(t)\rangle
+d_-(0)e^{i\theta_-(t)}e^{i\gamma_-(t)}|{-}(t)\rangle.
$
Next, we can expand the instantaneous eigenstates in the basis $|0\rangle, |1\rangle$ as
\be\label{uu}
|{\pm}(t)\rangle&=&u_{0\pm}(t)|0\rangle+u_{1\pm}(t)|1\rangle.
\ee
Therefore, the transforming matrix becomes
\be
U(t)=\begin{pmatrix}
|{+}(t)\rangle, |{-}(t)\rangle
\end{pmatrix}=\begin{pmatrix}
u_{0+}(t)&u_{0-}(t)\\
u_{1+}(t)&u_{1-}(t)
\end{pmatrix}.
\ee
Then, the coefficient $c_1(t)$ can be computed as
\be\label{c1t}
&&\hspace{-0.8cm}c_1(t)=\langle 1|\varphi(t)\rangle\nonumber\\
&&\hspace{-0.8cm}=\langle 1|d_+(0)e^{i\theta_+(t)}e^{i\gamma_+(t)}|{+}(t)\rangle+\langle 1|d_-(0)e^{i\theta_-(t)}e^{i\gamma_-(t)}|{-}(t)\rangle\nonumber\\
&&\hspace{-0.8cm}=d_+(0)u_{1+}(t)e^{i\theta_+(t)}e^{i\gamma_+(t)} +d_-(0)u_{1-}(t)e^{i\theta_-(t)}e^{i\gamma_-(t)}.\ \ 
\ee
Finally, Krylov complexity can be written in a closed form
\be\label{ktformula}
&&\hspace{-1cm}K(t)=|c_1(t)|^2
=|d_+(0)u_{1+}(t)|^2+|d_-(0)u_{1-}(t)|^2\nonumber\\
&&\hspace{-0.8cm}+2\text{Re}\left[d^*_+(0)u^*_{1+}(t)d_-(0)u_{1-}(t)e^{-i\Delta_\theta(t)}e^{-i\Delta_\gamma(t)}\right],
\ee
in which $\Delta_{\theta}(t)\equiv\theta_+(t)-\theta_-(t), \Delta_\gamma(t)\equiv\gamma_+(t)-\gamma_-(t)$. From this formula, it seems that the Krylov complexity has {\it explicit} dependence on the dynamical phase and Berry phase in the Krylov space. However, we will see that if we write Eq.\eqref{ktformula} explicitly in Lanczos coefficients, i.e., $a_0(t), a_1(t)$ and $b_1(t)$,  the Krylov complexity will {\it not} depend on the Berry phase $\gamma_\pm(t)$ in Krylov space! 

\paragraph{Explicit dependence of Krylov complexity on the Lanczos coefficients ---}
From the the effective Hamiltonian $H_{\text{eff}}$, 
its instantaneous eigenstates are 
\be
|{\pm}(t)\rangle=e^{i\phi_\pm(t)}|{\pm}(t)\rangle_R,
\ee
in which ${|{+}(t)\rangle_R\equiv(\cos\frac{\Theta(t)}{2}, \sin\frac{\Theta(t)}{2})^\text{T}}$, $|{-}(t)\rangle_R\equiv(-\sin\frac{\Theta(t)}{2}, \cos\frac{\Theta(t)}{2})^\text{T}$ and $\Theta(t)$ satisfies 
$
\cos\Theta(t)={\Delta(t)}/{\Omega(t)}$, ${\sin\Theta(t)={2b_1(t)}/{\Omega(t)}}
$. {The functions $\phi_\pm(t)$ and $\Theta(t)$ are all real functions. }
From Eq.\eqref{uu} it is easy to get
$
{u_{0+}(t)=e^{i\phi_+(t)}\cos\frac{\Theta}{2}}$, ${u_{1+}(t)=e^{i\phi_+(t)}\sin\frac{\Theta}{2}}$, 
${u_{0-}(t)=-e^{i\phi_-(t)}\sin\frac{\Theta}{2}}$, ${u_{1-}(t)=e^{i\phi_-(t)}\cos\frac{\Theta}{2}}.
$
From the relation ${2\cos^2\frac{\Theta}{2}=1+\cos\Theta}$ and ${2\sin^2\frac{\Theta}{2}=1-\cos\Theta}$, we can explicitly write down 
$
u_{1\pm}(t)=e^{i\phi_\pm(t)}\sqrt{\frac{1\mp\cos\Theta}{2}}=e^{i\phi_\pm(t)}\sqrt{\frac12\left(1\mp\frac{\Delta}{\Omega}\right)}.
$
From Eq.\eqref{d+d-} we get
$
d_+(0)=e^{-i\phi_+(0)}\cos\frac{\Theta(0)}{2}$ and $d_-(0)=-e^{-i\phi_-(0)}\sin\frac{\Theta(0)}{2}.
$
Then, the dynamical phase $\theta_\pm(t)$ becomes
$
\theta_\pm(t)=-\int_0^t\left(\frac{a_0(\tau)+a_1(\tau)}{2}\pm\frac{\Omega(\tau)}{2}\right)d\tau.
$
Thus,
$\Delta_\theta(t)=\theta_+(t)-\theta_-(t)=-\int_0^t\Omega(\tau)d\tau.$ 
 For the Berry phase, it is easy to get that $\!\ _R\langle{+}|\dot{+}\rangle_R=0$. Therefore,
$
\langle+(t)|\dot+(t)\rangle
=i\dot\phi_++\!\!\ _R\langle{+}|\dot{+}\rangle_R=i\dot\phi_+.$
Similarly, $\langle-|\dot-\rangle\!=\!i\dot\phi_-$. Thus, $\gamma_\pm(t)= -\int^t_0\dot\phi_\pm(\tau)d\tau =-\phi_\pm(t)+\phi_\pm(0)$ and $\Delta_\gamma=-\phi_+(t)+\phi_-(t)+\phi_+(0)-\phi_-(0)$. 
Finally, we get from Eq.\eqref{c1t} that
\be
c_1(t)\!&=&\!\cos\frac{\Theta(0)}{2}\sin\frac{\Theta(t)}{2}e^{-i\phi_+(0)}e^{i\phi_+(t)}e^{i\theta_+(t)}e^{i\gamma_+(t)}\nonumber\\
&&\hspace{-0.5cm}-\sin\frac{\Theta(0)}{2}\cos\frac{\Theta(t)}{2}e^{-i\phi_-(0)}e^{i\phi_-(t)}e^{i\theta_-(t)}e^{i\gamma_-(t)}\nonumber\\
&&\hspace{-1.5cm}=\cos\frac{\Theta(0)}{2}\sin\frac{\Theta(t)}{2}e^{i\theta_+(t)}-\sin\frac{\Theta(0)}{2}\cos\frac{\Theta(t)}{2}e^{i\theta_-(t)}.~~~~~
\ee
in which $\gamma_\pm(t)$ perfectly cancels with $e^{-i\phi_\pm(0)}e^{i\phi_\pm(t)}$. Finally,  we get the Krylov complexity as
\be
&&\hspace{-0.7cm}K(t)
=\bigg\lvert\cos\frac{\Theta(0)}{2}\sin\frac{\Theta(t)}{2}\bigg\rvert^2\!+\!\bigg\lvert\sin\frac{\Theta(0)}{2}\cos\frac{\Theta(t)}{2}\bigg\rvert^2\nonumber\\
&&\hspace{0.4cm}-2\cos\frac{\Theta(0)}{2}\sin\frac{\Theta(0)}{2}\sin\frac{\Theta(t)}{2}\cos\frac{\Theta(t)}{2}\cos\Delta_\theta(t)\nonumber\\
&=&\!\!\frac{1-\cos\Theta(0)\cos\Theta(t)}{2}-\frac12\sin\Theta(0)\sin\Theta(t)\cos\Delta_\theta(t)\nonumber\\
&=&\!\!\frac12\left(1-\frac{\Delta(0)\Delta(t)}{\Omega(0)\Omega(t)}\right)\!-\!\frac{2b_1(0)b_1(t)}{\Omega(0)\Omega(t)}\cos\left(\int_0^t\Omega(\tau)d\tau\right).~~~~~
\ee 
Obviously, if $a_0(t)=a_0, a_1(t)=a_1, b_1(t)=b_1$ are constants, the above formula of Krylov complexity returns to Eq.\eqref{Ktconst}. Therefore, in the quantum adiabatic dynamics, we find a closed form of the Krylov complexity expressed in Lanczos coefficients. Although the Berry phase in Krylov space, i.e. $\gamma_\pm(t)$ are absent in the final formula, the dynamical phase difference $\Delta_\theta=-\int_0^t\Omega(\tau)d\tau$ is retained. On the other hand, the information of Berry connection from the original Hilbert space still exists in Krylov complexity through the Lanczos coefficients and the quantity $A_m$. In general the integral $\int_0^t\Omega(\tau)d\tau$ may not be integrated to a closed form. However, as we noted above, if the coefficients are constants the integral will be analytically integrated and is linked to the Berry phase as we consider a closed evolving path. Therefore, we get the Krylov complexity controlled by the geometric phase.

\paragraph{Conclusions ---}
We studied the dynamics of the Krylov complexity in an adiabatic spin system and obtained a closed form of the Krylov complexity. It is found that the Krylov complexity depends on difference of the dynamical phase in the Krylov space, which is controlled by the Lanczos coefficients and further related to the Berry connections in the Hilbert space.  With the constant coefficients, we found that the frequency of the Krylov complexity was directly controlled by the Berry phase and the external field strength if the system evolved slowly in a closed path. Therefore, we established a bridge between the dynamics of complexity and the geometric phase in adiabatic approximations. 


Crucially, this geometric Berry phase shift of the frequency does not arise from non‑adiabatic level transitions. The system remains almost faithful to the instantaneous eigenstates, yet the superposition structure of the initial Krylov basis renders this geometric phase visible in the spreading dynamics of the Krylov subspace. In other words, even without level transitions to populate excited eigenstates via non‑adiabatic excitations, the geometry of the adiabatic path can still leave observable fingerprints on the Krylov complexity dynamics from the superposition states. This is a key insight of this letter.

This letter has bridged the Krylov complexity, a concept of spreading of states in quantum chaos, to the Berry phase, a geometric quantity in quantum dynamics.  This interesting connection will help us to understand the nature of chaos, integrability and geometry in quantum dynamics.  We expect that this connection can be readily extended to other models with geometric phases, such as the Aharonov-Bohm effect \cite{Aharonov:1959fk}. On the other hand, the connection between Krylov complexity and Berry phase provides us an alternative way to probe the geometric phase from Krylov complexity. 
\\

\acknowledgments
This work was supported by the National Natural Science Foundation of China (Grants No.12675057, No.12665009, No.1260051567); Hunan Natural Science Foundation (Grants No.2023JJ30487, No.2022JJ40340); Hunan Provincial Department of Education Project (Grant No.25B0480); The Key Project of Sichuan Science and Technology Education Joint Fund (Grant No.25LHJJ0097) and the Sichuan Natural Science Foundation (Grant No.2026NSFSC0746).




\begin{widetext}

\renewcommand{\theequation}{S\arabic{equation}}
\setcounter{equation}{0}

\begin{center}
\large{------ {\bf Supplemental Materials} ------}
\end{center}

\section{Derive the equation $i\partial_t|m(t)\rangle=-m\dot\varphi(1-\cos\theta)|m(t)\rangle$}
The instantaneous eigenstate basis $|m(t)\rangle$ can be rotated from the time-independent eigenstate $|m\rangle$ in fixed  $z$-basis as follows,
\be
|m(t)\rangle=\hat R(t)|m\rangle
\ee
where $\hat R(t)$ is a rotating operator defined as 
\be \label{euler}
\hat R(t)=e^{-i\varphi(t)\hat S_z}e^{-i\theta(t)\hat S_y}e^{+i\varphi(t)\hat S_z}
\ee
Therefore, $|\dot m(t)\rangle=\dot{\hat{R}}(t)|m\rangle$ with 
\be
\dot{\hat{R}}(t)&=&-i\dot\varphi\hat S_ze^{-i\varphi(t)\hat S_z}e^{-i\theta(t)\hat S_y}e^{+i\varphi(t)\hat S_z}+e^{-i\varphi(t)\hat S_z}(-i\dot\theta\hat S_y)e^{-i\theta(t)\hat S_y}e^{+i\varphi(t)\hat S_z}\nonumber\\
&&+e^{-i\varphi(t)\hat S_z}e^{-i\theta(t)\hat S_y}(i\dot\varphi\hat S_z)e^{+i\varphi(t)\hat S_z}.
\ee
Then $\hat R^\dagger\dot{\hat{R}}$ becomes
\be
\hat R^\dagger\dot{\hat{R}}&=&e^{-i\varphi(t)\hat S_z}e^{+i\theta(t)\hat S_y}e^{+i\varphi(t)\hat S_z}
\big(-i\dot\varphi\hat S_ze^{-i\varphi(t)\hat S_z}e^{-i\theta(t)\hat S_y}e^{+i\varphi(t)\hat S_z}+e^{-i\varphi(t)\hat S_z}(-i\dot\theta\hat S_y)e^{-i\theta(t)\hat S_y}e^{+i\varphi(t)\hat S_z}\nonumber\\
&&\hspace{4cm}+e^{-i\varphi(t)\hat S_z}e^{-i\theta(t)\hat S_y}(i\dot\varphi\hat S_z)e^{+i\varphi(t)\hat S_z}\big)\nonumber\\
&=&e^{-i\varphi(t)\hat S_z}e^{+i\theta(t)\hat S_y}(-i\dot\varphi\hat S_z)e^{-i\theta(t)\hat S_y}e^{+i\varphi(t)\hat S_z}+e^{-i\varphi(t)\hat S_z}e^{+i\theta(t)\hat S_y}(-i\dot\theta\hat S_y)e^{-i\theta(t)\hat S_y}e^{+i\varphi(t)\hat S_z}\nonumber\\
&&+e^{-i\varphi(t)\hat S_z}(i\dot\varphi\hat S_z)e^{+i\varphi(t)\hat S_z}
\ee
Using the Baker-Campbell-Hausdorff (BCH) formula
\be
e^{\hat A}\hat B e^{-\hat A}=\sum_{n=0}^\infty\frac{1}{n!}[\hat A^{(n)},\hat B],
\ee
where $[\hat A^{(n)},\hat B]\equiv\underbrace{[\hat A,[\hat A,\cdots,[\hat A}_n,B]$, it is easy to get that 
\be\label{S6}
e^{-i\varphi\hat S_z}\hat S_y e^{+i\varphi\hat S_z}&=&\sum_{n=0}^\infty\frac{1}{n!}[(-i\varphi\hat S_z)^{(n)},\hat S_y]=\sum_{n=0}^\infty \frac{(-i\varphi)^n}{n!}[\hat S_z^{(n)},\hat S_y]\nonumber\\
&=&\hat S_y-\varphi\hat S_x-\frac{\varphi^2}{2!}\hat S_y+\frac{\varphi^3}{3!}\hat S_x+\frac{\varphi^4}{4!}\hat S_y-\frac{\varphi^5}{5!}\hat S_x+\cdots\nonumber\\
&=&\left(1-\frac{\varphi^2}{2!}+\frac{\varphi^4}{4!}-\cdots\right)\hat S_y+\left(-\varphi+\frac{\varphi^3}{3!}-\frac{\varphi^5}{5!}+\cdots\right)\hat S_x\nonumber\\
&=&\cos\varphi\hat S_y-\sin\varphi\hat S_x.
\ee
Similarly, $e^{+i\theta(t)\hat S_y}\hat S_z e^{-i\theta(t)\hat S_y}=\cos\theta\hat S_z-\sin\theta\hat S_x$ and $e^{-i\varphi\hat S_z}\hat S_x e^{i\varphi\hat S_z}=\cos\theta\hat S_x+\sin\varphi\hat S_y$. 

Therefore,
\be
\hat R^\dagger\dot{\hat{R}}&=&-i\dot\varphi e^{-i\varphi(t)\hat S_z}e^{+i\theta(t)\hat S_y}\hat S_z e^{-i\theta(t)\hat S_y}e^{+i\varphi(t)\hat S_z} -i\dot\theta e^{-i\varphi(t)\hat S_z}\hat S_y e^{+i\varphi(t)\hat S_z}+i\dot\varphi\hat S_z\nonumber\\
&=&-i\dot\varphi e^{-i\varphi\hat S_z}(\cos\theta\hat S_z-\sin\theta\hat S_x)e^{i\varphi\hat S_z}-i\dot\theta(-\sin\varphi\hat S_x+\cos\varphi\hat S_y)+i\dot\varphi\hat S_z\nonumber\\
&=&-i\dot\varphi\cos\theta\hat S_z+i\dot\varphi\sin\theta e^{-i\varphi\hat S_z}\hat S_x e^{i\varphi\hat S_z}-i\dot\theta(-\sin\varphi\hat S_x+\cos\varphi\hat S_y)+i\dot\varphi\hat S_z\nonumber\\
&=&-i\dot\varphi\cos\theta\hat S_z+i\dot\varphi\sin\theta(\cos\theta\hat S_x+\sin\varphi\hat S_y)-i\dot\theta(-\sin\varphi\hat S_x+\cos\varphi\hat S_y)+i\dot\varphi\hat S_z.
\ee
Because of $\langle m|\hat S_x|m\rangle=\langle m|\hat S_y|m\rangle=0$, thus
\be
i\langle m(t)|\dot m(t)\rangle=i\langle m|\hat R^\dagger\dot{\hat{R}}|m\rangle=i\big(-i\dot\varphi\cos\theta \langle m|\hat S_z|m\rangle+i\dot\varphi\langle m|\hat S_z|m\rangle\big)=-m\dot\varphi(1-\cos\theta).
\ee
So, the original equation $i\partial_t|m(t)\rangle=|m(t)\rangle i\langle m(t)|\dot m(t)\rangle$ becomes $i\partial_t|m(t)\rangle=-m\dot\varphi(1-\cos\theta)|m(t)\rangle$. \\

{\bf Note added:} Our definition of $\hat R(t)$ in Eq.\eqref{euler} is equivalent to the definition of $\hat R(t)$ in \cite{Takahashi:2024hex} that 
\be
\hat R(t)\equiv e^{-i\theta(t) \hat {\bf S}\cdot {\bf e}_\varphi(t)},
\ee
where ${\bf e}_\varphi(t)=(-\sin\varphi(t),\cos\varphi(t),0)$.
In fact we know the following identity
\be
e^{-i\varphi(t)\hat S_z}e^{-i\theta(t)\hat S_y}e^{+i\varphi(t)\hat S_z}=\exp(e^{-i\varphi\hat S_z}(-i\theta\hat S_y)e^{+i\varphi\hat S_z})
\ee
The proof is as follows:
\be
&&e^{-i\varphi\hat S_z}e^{-i\theta\hat S_y}e^{+i\varphi\hat S_z}=e^{-i\varphi\hat S_z}\sum_{n=0}^\infty\frac{(-i\theta\hat S_y)^n}{n!}e^{+i\varphi\hat S_z}=\sum_{n=0}^\infty\frac{1}{n!}e^{-i\varphi\hat S_z}(-i\theta\hat S_y)^ne^{+i\varphi\hat S_z}\nonumber\\
&=&\sum_{n=0}^\infty\frac{1}{n!}\underbrace{e^{-i\varphi\hat S_z}(-i\theta\hat S_y)e^{+i\varphi\hat S_z}e^{-i\varphi\hat S_z}(-i\theta\hat S_y)e^{+i\varphi\hat S_z}\cdots e^{-i\varphi\hat S_z}(-i\theta\hat S_y)e^{+i\varphi\hat S_z}}_n\nonumber\\
&=&\sum_{n=0}^\infty\frac{1}{n!}\left(e^{-i\varphi\hat S_z}(-i\theta\hat S_y)e^{+i\varphi\hat S_z}\right)^n\nonumber\\
&=&\exp(e^{-i\varphi\hat S_z}(-i\theta\hat S_y)e^{+i\varphi\hat S_z}).
\ee
Then,  from Eq.\eqref{S6} we get
\be
\exp(e^{-i\varphi\hat S_z}(-i\theta\hat S_y)e^{+i\varphi\hat S_z})=\exp\left(-i\theta\left(\cos\varphi\hat S_y-\sin\varphi\hat S_x\right)\right)=\exp\left(-i\theta\hat {\bf S}\cdot{\bf e}_\varphi\right)
\ee
Therefore, we complete the proof. \hfill$\blacksquare$

\section{Krylov complexity with constant coefficients}
When the Lanczos coefficients $a_0(t)=a_0, a_1(t)=a_1, b_1(t)=b_1$ are constants, we can directly solve the EoM
\be\label{scheom}
i\partial_t|\varphi(t)\rangle=\mathcal{\hat L}|\varphi(t)\rangle
\ee
where $|\varphi(t)\rangle=(c_0(t),c_1(t))^T$. The initial condition is $|\varphi(0)\rangle={{c_0(0)}\choose{c_1(0)}}={1\choose 0}$. Therefore, the equation \eqref{scheom} becomes 
\be
i\begin{pmatrix}
\dot c_0(t)\\
\dot c_1(t)
\end{pmatrix}=\begin{pmatrix}
a_0 &b_1 \\
b_1 &a_1
\end{pmatrix}\begin{pmatrix}
c_0(t)\\
c_1(t)
\end{pmatrix}
\ee
With these conditions we can directly solve from the \texttt{DSolve} command in {\it Wolfram Mathematica} that 
\be
c_1(t)=-\frac{2 i {b_1} e^{-\frac{i}{2} t ({a_0}+{a_1})} \sin\left(\frac{1}{2} t \sqrt{({a_0}-{a_1})^2+4{b_1}^2}\right)}{\sqrt{({a_0}-{a_1})^2+4{b_1}^2}}
\ee
Therefore, 
\be\label{Ktconst}
K(t)=|c_1(t)|^2=\frac{4b_1^2\sin ^2\left(\frac{1}{2} t \sqrt{(a_0-a_1)^2+4b_1^2}\right)}{(a_0-a_1)^2+4b_1^2}
\ee

\begin{figure}[t]
\centering
\includegraphics[trim=0.cm 0.cm 0cm 0cm, clip=true, scale=0.55]{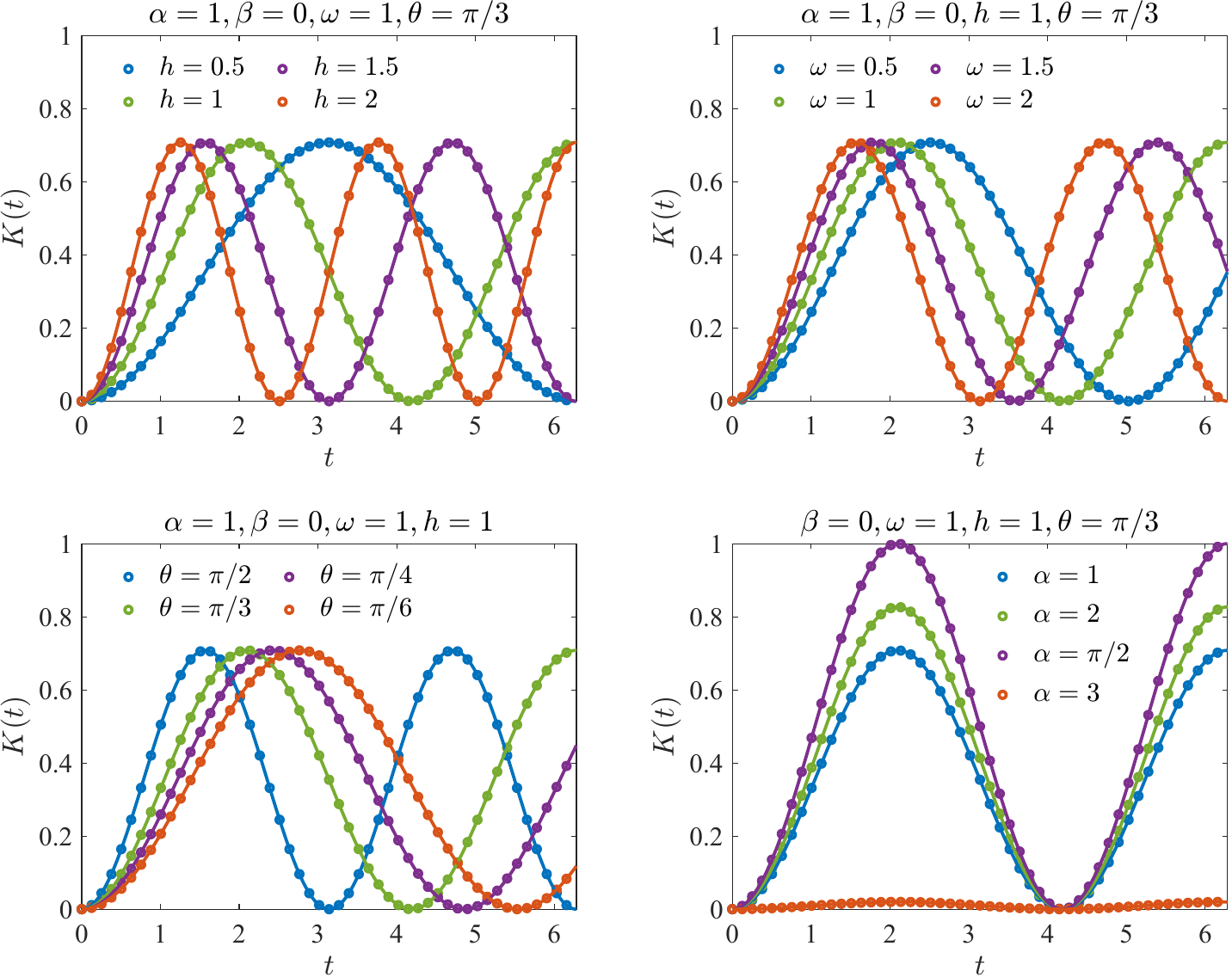}
\caption{Time evolution of Krylov complexity with various constant parameters in a single qubit system. In this case, they perform harmonic oscillation behaviors. Numerical results (colorful circles) match theoretical predictions from Eq.\eqref{Ktconst} (solid lines) very well. In these panels, $\beta\equiv\dot\varphi$ is the constant angular velocity.}\label{p2}
\end{figure}

With the constants, the Lanczos coefficients can be explicitly expressed as
\be
\begin{cases}
|K_0(t)\rangle=\cos\left(\frac{\alpha}{2}\right)|{+}(t)\rangle+\sin\left(\frac{\alpha}{2}\right)e^{i\beta}|{-}(t)\rangle,\\
a_0=\frac12 \left(h+\dot\varphi(1-\cos\theta)\right)\cos\alpha,\\
b_0=0,
\end{cases}
\begin{cases}
|K_1(t)\rangle=\sin\left(\frac{\alpha}{2}\right)|{+}(t)\rangle-\cos\left(\frac{\alpha}{2}\right)e^{i\beta}|{-}(t)\rangle,\\
a_1=-\frac12 \left(h+\dot\varphi(1-\cos\theta)\right)\cos\alpha,\\
b_1=\frac{1}{2}\left(h+\dot\varphi(1-\cos\theta)\right)\sin\alpha.
\end{cases}
\ee
Therefore, $a_0-a_1=\left(h+\dot\varphi(1-\cos\theta)\right)\cos\alpha$, $(a_0-a_1)^2+4b_1^2=\left(h+\dot\varphi(1-\cos\theta)\right)^2$. Hence, we can further express Eq.\eqref{Ktconst} as
\be\label{s18}
K(t)=\sin^2\alpha\,\sin^2\left(\frac{h+\dot\varphi(1-\cos\theta)}{2}t\right),
\ee
which is exactly the Eq.(8) in the main text.

In Fig.\ref{p2} we show the numerical results (colorful circles) and theoretical predictions from Eq.\eqref{Ktconst} (solid lines) with various constant parameters. They match each other very well. It is found that in this case they perform harmonic oscillations in time with the frequencies exactly identical to Eq.\eqref{s18}. 

\end{widetext}

\bibliography{refs.bib}

\begin{thebibliography}{56}%
\makeatletter
\providecommand \@ifxundefined [1]{%
 \@ifx{#1\undefined}
}%
\providecommand \@ifnum [1]{%
 \ifnum #1\expandafter \@firstoftwo
 \else \expandafter \@secondoftwo
 \fi
}%
\providecommand \@ifx [1]{%
 \ifx #1\expandafter \@firstoftwo
 \else \expandafter \@secondoftwo
 \fi
}%
\providecommand \natexlab [1]{#1}%
\providecommand \enquote  [1]{``#1''}%
\providecommand \bibnamefont  [1]{#1}%
\providecommand \bibfnamefont [1]{#1}%
\providecommand \citenamefont [1]{#1}%
\providecommand \href@noop [0]{\@secondoftwo}%
\providecommand \href [0]{\begingroup \@sanitize@url \@href}%
\providecommand \@href[1]{\@@startlink{#1}\@@href}%
\providecommand \@@href[1]{\endgroup#1\@@endlink}%
\providecommand \@sanitize@url [0]{\catcode `\\12\catcode `\$12\catcode
  `\&12\catcode `\#12\catcode `\^12\catcode `\_12\catcode `\%12\relax}%
\providecommand \@@startlink[1]{}%
\providecommand \@@endlink[0]{}%
\providecommand \url  [0]{\begingroup\@sanitize@url \@url }%
\providecommand \@url [1]{\endgroup\@href {#1}{\urlprefix }}%
\providecommand \urlprefix  [0]{URL }%
\providecommand \Eprint [0]{\href }%
\providecommand \doibase [0]{https://doi.org/}%
\providecommand \selectlanguage [0]{\@gobble}%
\providecommand \bibinfo  [0]{\@secondoftwo}%
\providecommand \bibfield  [0]{\@secondoftwo}%
\providecommand \translation [1]{[#1]}%
\providecommand \BibitemOpen [0]{}%
\providecommand \bibitemStop [0]{}%
\providecommand \bibitemNoStop [0]{.\EOS\space}%
\providecommand \EOS [0]{\spacefactor3000\relax}%
\providecommand \BibitemShut  [1]{\csname bibitem#1\endcsname}%
\let\auto@bib@innerbib\@empty
\bibitem [{\citenamefont {Chapman}\ and\ \citenamefont
  {Policastro}(2022)}]{Chapman:2021jbh}%
  \BibitemOpen
  \bibfield  {author} {\bibinfo {author} {\bibfnamefont {S.}~\bibnamefont
  {Chapman}}\ and\ \bibinfo {author} {\bibfnamefont {G.}~\bibnamefont
  {Policastro}},\ }\bibfield  {title} {\bibinfo {title} {{Quantum computational
  complexity from quantum information to black holes and back}},\ }\href
  {https://doi.org/10.1140/epjc/s10052-022-10037-1} {\bibfield  {journal}
  {\bibinfo  {journal} {Eur. Phys. J. C}\ }\textbf {\bibinfo {volume} {82}},\
  \bibinfo {pages} {128} (\bibinfo {year} {2022})},\ \Eprint
  {https://arxiv.org/abs/2110.14672} {arXiv:2110.14672 [hep-th]} \BibitemShut
  {NoStop}%
\bibitem [{\citenamefont {Baiguera}\ \emph {et~al.}(2026)\citenamefont
  {Baiguera}, \citenamefont {Balasubramanian}, \citenamefont {Caputa},
  \citenamefont {Chapman}, \citenamefont {Haferkamp}, \citenamefont {Heller},\
  and\ \citenamefont {Halpern}}]{Baiguera:2025dkc}%
  \BibitemOpen
  \bibfield  {author} {\bibinfo {author} {\bibfnamefont {S.}~\bibnamefont
  {Baiguera}}, \bibinfo {author} {\bibfnamefont {V.}~\bibnamefont
  {Balasubramanian}}, \bibinfo {author} {\bibfnamefont {P.}~\bibnamefont
  {Caputa}}, \bibinfo {author} {\bibfnamefont {S.}~\bibnamefont {Chapman}},
  \bibinfo {author} {\bibfnamefont {J.}~\bibnamefont {Haferkamp}}, \bibinfo
  {author} {\bibfnamefont {M.~P.}\ \bibnamefont {Heller}},\ and\ \bibinfo
  {author} {\bibfnamefont {N.~Y.}\ \bibnamefont {Halpern}},\ }\bibfield
  {title} {\bibinfo {title} {{Quantum complexity in gravity, quantum field
  theory, and quantum information science}},\ }\href
  {https://doi.org/10.1016/j.physrep.2025.11.001} {\bibfield  {journal}
  {\bibinfo  {journal} {Phys. Rept.}\ }\textbf {\bibinfo {volume} {1159}},\
  \bibinfo {pages} {1} (\bibinfo {year} {2026})},\ \Eprint
  {https://arxiv.org/abs/2503.10753} {arXiv:2503.10753 [hep-th]} \BibitemShut
  {NoStop}%
\bibitem [{\citenamefont {Nandy}\ \emph {et~al.}(2025)\citenamefont {Nandy},
  \citenamefont {Matsoukas-Roubeas}, \citenamefont {Mart{\'\i}nez-Azcona},
  \citenamefont {Dymarsky},\ and\ \citenamefont {del Campo}}]{Nandy:2024evd}%
  \BibitemOpen
  \bibfield  {author} {\bibinfo {author} {\bibfnamefont {P.}~\bibnamefont
  {Nandy}}, \bibinfo {author} {\bibfnamefont {A.~S.}\ \bibnamefont
  {Matsoukas-Roubeas}}, \bibinfo {author} {\bibfnamefont {P.}~\bibnamefont
  {Mart{\'\i}nez-Azcona}}, \bibinfo {author} {\bibfnamefont {A.}~\bibnamefont
  {Dymarsky}},\ and\ \bibinfo {author} {\bibfnamefont {A.}~\bibnamefont {del
  Campo}},\ }\bibfield  {title} {\bibinfo {title} {{Quantum dynamics in Krylov
  space: Methods and applications}},\ }\href
  {https://doi.org/10.1016/j.physrep.2025.05.001} {\bibfield  {journal}
  {\bibinfo  {journal} {Phys. Rept.}\ }\textbf {\bibinfo {volume}
  {1125-1128}},\ \bibinfo {pages} {1} (\bibinfo {year} {2025})},\ \Eprint
  {https://arxiv.org/abs/2405.09628} {arXiv:2405.09628 [quant-ph]} \BibitemShut
  {NoStop}%
\bibitem [{\citenamefont {Rabinovici}\ \emph {et~al.}(2025)\citenamefont
  {Rabinovici}, \citenamefont {S{\'a}nchez-Garrido}, \citenamefont {Shir},\
  and\ \citenamefont {Sonner}}]{Rabinovici:2025otw}%
  \BibitemOpen
  \bibfield  {author} {\bibinfo {author} {\bibfnamefont {E.}~\bibnamefont
  {Rabinovici}}, \bibinfo {author} {\bibfnamefont {A.}~\bibnamefont
  {S{\'a}nchez-Garrido}}, \bibinfo {author} {\bibfnamefont {R.}~\bibnamefont
  {Shir}},\ and\ \bibinfo {author} {\bibfnamefont {J.}~\bibnamefont {Sonner}},\
  }\bibfield  {title} {\bibinfo {title} {{Krylov Complexity}},\ }\href@noop {}
  {\  (\bibinfo {year} {2025})},\ \Eprint {https://arxiv.org/abs/2507.06286}
  {arXiv:2507.06286 [hep-th]} \BibitemShut {NoStop}%
\bibitem [{\citenamefont {Parker}\ \emph {et~al.}(2019)\citenamefont {Parker},
  \citenamefont {Cao}, \citenamefont {Avdoshkin}, \citenamefont {Scaffidi},\
  and\ \citenamefont {Altman}}]{Parker:2018yvk}%
  \BibitemOpen
  \bibfield  {author} {\bibinfo {author} {\bibfnamefont {D.~E.}\ \bibnamefont
  {Parker}}, \bibinfo {author} {\bibfnamefont {X.}~\bibnamefont {Cao}},
  \bibinfo {author} {\bibfnamefont {A.}~\bibnamefont {Avdoshkin}}, \bibinfo
  {author} {\bibfnamefont {T.}~\bibnamefont {Scaffidi}},\ and\ \bibinfo
  {author} {\bibfnamefont {E.}~\bibnamefont {Altman}},\ }\bibfield  {title}
  {\bibinfo {title} {{A Universal Operator Growth Hypothesis}},\ }\href
  {https://doi.org/10.1103/PhysRevX.9.041017} {\bibfield  {journal} {\bibinfo
  {journal} {Phys. Rev. X}\ }\textbf {\bibinfo {volume} {9}},\ \bibinfo {pages}
  {041017} (\bibinfo {year} {2019})},\ \Eprint
  {https://arxiv.org/abs/1812.08657} {arXiv:1812.08657 [cond-mat.stat-mech]}
  \BibitemShut {NoStop}%
\bibitem [{\citenamefont {Balasubramanian}\ \emph {et~al.}(2022)\citenamefont
  {Balasubramanian}, \citenamefont {Caputa}, \citenamefont {Magan},\ and\
  \citenamefont {Wu}}]{Balasubramanian:2022tpr}%
  \BibitemOpen
  \bibfield  {author} {\bibinfo {author} {\bibfnamefont {V.}~\bibnamefont
  {Balasubramanian}}, \bibinfo {author} {\bibfnamefont {P.}~\bibnamefont
  {Caputa}}, \bibinfo {author} {\bibfnamefont {J.~M.}\ \bibnamefont {Magan}},\
  and\ \bibinfo {author} {\bibfnamefont {Q.}~\bibnamefont {Wu}},\ }\bibfield
  {title} {\bibinfo {title} {{Quantum chaos and the complexity of spread of
  states}},\ }\href {https://doi.org/10.1103/PhysRevD.106.046007} {\bibfield
  {journal} {\bibinfo  {journal} {Phys. Rev. D}\ }\textbf {\bibinfo {volume}
  {106}},\ \bibinfo {pages} {046007} (\bibinfo {year} {2022})},\ \Eprint
  {https://arxiv.org/abs/2202.06957} {arXiv:2202.06957 [hep-th]} \BibitemShut
  {NoStop}%
\bibitem [{\citenamefont {Rabinovici}\ \emph {et~al.}(2023)\citenamefont
  {Rabinovici}, \citenamefont {S{\'a}nchez-Garrido}, \citenamefont {Shir},\
  and\ \citenamefont {Sonner}}]{Rabinovici:2023yex}%
  \BibitemOpen
  \bibfield  {author} {\bibinfo {author} {\bibfnamefont {E.}~\bibnamefont
  {Rabinovici}}, \bibinfo {author} {\bibfnamefont {A.}~\bibnamefont
  {S{\'a}nchez-Garrido}}, \bibinfo {author} {\bibfnamefont {R.}~\bibnamefont
  {Shir}},\ and\ \bibinfo {author} {\bibfnamefont {J.}~\bibnamefont {Sonner}},\
  }\bibfield  {title} {\bibinfo {title} {{A bulk manifestation of Krylov
  complexity}},\ }\href {https://doi.org/10.1007/JHEP08(2023)213} {\bibfield
  {journal} {\bibinfo  {journal} {JHEP}\ }\textbf {\bibinfo {volume} {08}},\
  \bibinfo {pages} {213}},\ \Eprint {https://arxiv.org/abs/2305.04355}
  {arXiv:2305.04355 [hep-th]} \BibitemShut {NoStop}%
\bibitem [{\citenamefont {Ambrosini}\ \emph {et~al.}(2025)\citenamefont
  {Ambrosini}, \citenamefont {Rabinovici}, \citenamefont {S{\'a}nchez-Garrido},
  \citenamefont {Shir},\ and\ \citenamefont {Sonner}}]{Ambrosini:2024sre}%
  \BibitemOpen
  \bibfield  {author} {\bibinfo {author} {\bibfnamefont {M.}~\bibnamefont
  {Ambrosini}}, \bibinfo {author} {\bibfnamefont {E.}~\bibnamefont
  {Rabinovici}}, \bibinfo {author} {\bibfnamefont {A.}~\bibnamefont
  {S{\'a}nchez-Garrido}}, \bibinfo {author} {\bibfnamefont {R.}~\bibnamefont
  {Shir}},\ and\ \bibinfo {author} {\bibfnamefont {J.}~\bibnamefont {Sonner}},\
  }\bibfield  {title} {\bibinfo {title} {{Operator K-complexity in DSSYK:
  Krylov complexity equals bulk length}},\ }\href
  {https://doi.org/10.1007/JHEP08(2025)059} {\bibfield  {journal} {\bibinfo
  {journal} {JHEP}\ }\textbf {\bibinfo {volume} {08}},\ \bibinfo {pages}
  {059}},\ \Eprint {https://arxiv.org/abs/2412.15318} {arXiv:2412.15318
  [hep-th]} \BibitemShut {NoStop}%
\bibitem [{\citenamefont {Lin}(2022)}]{Lin:2022rbf}%
  \BibitemOpen
  \bibfield  {author} {\bibinfo {author} {\bibfnamefont {H.~W.}\ \bibnamefont
  {Lin}},\ }\bibfield  {title} {\bibinfo {title} {{The bulk Hilbert space of
  double scaled SYK}},\ }\href {https://doi.org/10.1007/JHEP11(2022)060}
  {\bibfield  {journal} {\bibinfo  {journal} {JHEP}\ }\textbf {\bibinfo
  {volume} {11}},\ \bibinfo {pages} {060}},\ \Eprint
  {https://arxiv.org/abs/2208.07032} {arXiv:2208.07032 [hep-th]} \BibitemShut
  {NoStop}%
\bibitem [{\citenamefont {Dymarsky}\ and\ \citenamefont
  {Smolkin}(2021)}]{Dymarsky:2021bjq}%
  \BibitemOpen
  \bibfield  {author} {\bibinfo {author} {\bibfnamefont {A.}~\bibnamefont
  {Dymarsky}}\ and\ \bibinfo {author} {\bibfnamefont {M.}~\bibnamefont
  {Smolkin}},\ }\bibfield  {title} {\bibinfo {title} {{Krylov complexity in
  conformal field theory}},\ }\href
  {https://doi.org/10.1103/PhysRevD.104.L081702} {\bibfield  {journal}
  {\bibinfo  {journal} {Phys. Rev. D}\ }\textbf {\bibinfo {volume} {104}},\
  \bibinfo {pages} {L081702} (\bibinfo {year} {2021})},\ \Eprint
  {https://arxiv.org/abs/2104.09514} {arXiv:2104.09514 [hep-th]} \BibitemShut
  {NoStop}%
\bibitem [{\citenamefont {Kundu}\ \emph {et~al.}(2023)\citenamefont {Kundu},
  \citenamefont {Malvimat},\ and\ \citenamefont {Sinha}}]{Kundu:2023hbk}%
  \BibitemOpen
  \bibfield  {author} {\bibinfo {author} {\bibfnamefont {A.}~\bibnamefont
  {Kundu}}, \bibinfo {author} {\bibfnamefont {V.}~\bibnamefont {Malvimat}},\
  and\ \bibinfo {author} {\bibfnamefont {R.}~\bibnamefont {Sinha}},\ }\bibfield
   {title} {\bibinfo {title} {{State dependence of Krylov complexity in 2d
  CFTs}},\ }\href {https://doi.org/10.1007/JHEP09(2023)011} {\bibfield
  {journal} {\bibinfo  {journal} {JHEP}\ }\textbf {\bibinfo {volume} {09}},\
  \bibinfo {pages} {011}},\ \Eprint {https://arxiv.org/abs/2303.03426}
  {arXiv:2303.03426 [hep-th]} \BibitemShut {NoStop}%
\bibitem [{\citenamefont {Dymarsky}\ and\ \citenamefont
  {Gorsky}(2020)}]{Dymarsky:2019elm}%
  \BibitemOpen
  \bibfield  {author} {\bibinfo {author} {\bibfnamefont {A.}~\bibnamefont
  {Dymarsky}}\ and\ \bibinfo {author} {\bibfnamefont {A.}~\bibnamefont
  {Gorsky}},\ }\bibfield  {title} {\bibinfo {title} {{Quantum chaos as
  delocalization in Krylov space}},\ }\href
  {https://doi.org/10.1103/PhysRevB.102.085137} {\bibfield  {journal} {\bibinfo
   {journal} {Phys. Rev. B}\ }\textbf {\bibinfo {volume} {102}},\ \bibinfo
  {pages} {085137} (\bibinfo {year} {2020})},\ \Eprint
  {https://arxiv.org/abs/1912.12227} {arXiv:1912.12227 [cond-mat.stat-mech]}
  \BibitemShut {NoStop}%
\bibitem [{\citenamefont {Trigueros}\ and\ \citenamefont
  {Lin}(2022)}]{Trigueros:2021rwj}%
  \BibitemOpen
  \bibfield  {author} {\bibinfo {author} {\bibfnamefont {F.~B.}\ \bibnamefont
  {Trigueros}}\ and\ \bibinfo {author} {\bibfnamefont {C.-J.}\ \bibnamefont
  {Lin}},\ }\bibfield  {title} {\bibinfo {title} {{Krylov complexity of
  many-body localization: Operator localization in Krylov basis}},\ }\href
  {https://doi.org/10.21468/SciPostPhys.13.2.037} {\bibfield  {journal}
  {\bibinfo  {journal} {SciPost Phys.}\ }\textbf {\bibinfo {volume} {13}},\
  \bibinfo {pages} {037} (\bibinfo {year} {2022})},\ \Eprint
  {https://arxiv.org/abs/2112.04722} {arXiv:2112.04722 [cond-mat.dis-nn]}
  \BibitemShut {NoStop}%
\bibitem [{\citenamefont {Rabinovici}\ \emph
  {et~al.}(2022{\natexlab{a}})\citenamefont {Rabinovici}, \citenamefont
  {S{\'a}nchez-Garrido}, \citenamefont {Shir},\ and\ \citenamefont
  {Sonner}}]{Rabinovici:2021qqt}%
  \BibitemOpen
  \bibfield  {author} {\bibinfo {author} {\bibfnamefont {E.}~\bibnamefont
  {Rabinovici}}, \bibinfo {author} {\bibfnamefont {A.}~\bibnamefont
  {S{\'a}nchez-Garrido}}, \bibinfo {author} {\bibfnamefont {R.}~\bibnamefont
  {Shir}},\ and\ \bibinfo {author} {\bibfnamefont {J.}~\bibnamefont {Sonner}},\
  }\bibfield  {title} {\bibinfo {title} {{Krylov localization and suppression
  of complexity}},\ }\href {https://doi.org/10.1007/JHEP03(2022)211} {\bibfield
   {journal} {\bibinfo  {journal} {JHEP}\ }\textbf {\bibinfo {volume} {03}},\
  \bibinfo {pages} {211}},\ \Eprint {https://arxiv.org/abs/2112.12128}
  {arXiv:2112.12128 [hep-th]} \BibitemShut {NoStop}%
\bibitem [{\citenamefont {Rabinovici}\ \emph
  {et~al.}(2022{\natexlab{b}})\citenamefont {Rabinovici}, \citenamefont
  {S{\'a}nchez-Garrido}, \citenamefont {Shir},\ and\ \citenamefont
  {Sonner}}]{Rabinovici:2022beu}%
  \BibitemOpen
  \bibfield  {author} {\bibinfo {author} {\bibfnamefont {E.}~\bibnamefont
  {Rabinovici}}, \bibinfo {author} {\bibfnamefont {A.}~\bibnamefont
  {S{\'a}nchez-Garrido}}, \bibinfo {author} {\bibfnamefont {R.}~\bibnamefont
  {Shir}},\ and\ \bibinfo {author} {\bibfnamefont {J.}~\bibnamefont {Sonner}},\
  }\bibfield  {title} {\bibinfo {title} {{Krylov complexity from integrability
  to chaos}},\ }\href {https://doi.org/10.1007/JHEP07(2022)151} {\bibfield
  {journal} {\bibinfo  {journal} {JHEP}\ }\textbf {\bibinfo {volume} {07}},\
  \bibinfo {pages} {151}},\ \Eprint {https://arxiv.org/abs/2207.07701}
  {arXiv:2207.07701 [hep-th]} \BibitemShut {NoStop}%
\bibitem [{\citenamefont {Caputa}\ \emph {et~al.}(2022)\citenamefont {Caputa},
  \citenamefont {Magan},\ and\ \citenamefont {Patramanis}}]{Caputa:2021sib}%
  \BibitemOpen
  \bibfield  {author} {\bibinfo {author} {\bibfnamefont {P.}~\bibnamefont
  {Caputa}}, \bibinfo {author} {\bibfnamefont {J.~M.}\ \bibnamefont {Magan}},\
  and\ \bibinfo {author} {\bibfnamefont {D.}~\bibnamefont {Patramanis}},\
  }\bibfield  {title} {\bibinfo {title} {{Geometry of Krylov complexity}},\
  }\href {https://doi.org/10.1103/PhysRevResearch.4.013041} {\bibfield
  {journal} {\bibinfo  {journal} {Phys. Rev. Res.}\ }\textbf {\bibinfo {volume}
  {4}},\ \bibinfo {pages} {013041} (\bibinfo {year} {2022})},\ \Eprint
  {https://arxiv.org/abs/2109.03824} {arXiv:2109.03824 [hep-th]} \BibitemShut
  {NoStop}%
\bibitem [{\citenamefont {Zhai}\ \emph {et~al.}(2026)\citenamefont {Zhai},
  \citenamefont {Liu},\ and\ \citenamefont {Zhang}}]{Zhai:2024tkz}%
  \BibitemOpen
  \bibfield  {author} {\bibinfo {author} {\bibfnamefont {K.-H.}\ \bibnamefont
  {Zhai}}, \bibinfo {author} {\bibfnamefont {L.-H.}\ \bibnamefont {Liu}},\ and\
  \bibinfo {author} {\bibfnamefont {H.-Q.}\ \bibnamefont {Zhang}},\ }\bibfield
  {title} {\bibinfo {title} {{Generalized CV Conjecture and Krylov Complexity
  in Two-Mode Hermitian Systems via Information Geometry}},\ }\href
  {https://doi.org/10.1016/j.aop.2026.170534} {\bibfield  {journal} {\bibinfo
  {journal} {Annals Phys.}\ }\textbf {\bibinfo {volume} {491}},\ \bibinfo
  {pages} {170534} (\bibinfo {year} {2026})},\ \Eprint
  {https://arxiv.org/abs/2412.08925} {arXiv:2412.08925 [hep-th]} \BibitemShut
  {NoStop}%
\bibitem [{\citenamefont {Grabarits}\ \emph {et~al.}(2026)\citenamefont
  {Grabarits}, \citenamefont {Medina-Guerra},\ and\ \citenamefont {del
  Campo}}]{Grabarits:2026hjz}%
  \BibitemOpen
  \bibfield  {author} {\bibinfo {author} {\bibfnamefont {A.}~\bibnamefont
  {Grabarits}}, \bibinfo {author} {\bibfnamefont {E.}~\bibnamefont
  {Medina-Guerra}},\ and\ \bibinfo {author} {\bibfnamefont {A.}~\bibnamefont
  {del Campo}},\ }\bibfield  {title} {\bibinfo {title} {{Krylov Dynamics and
  Operator Growth in Time-Dependent Systems via Lie Algebras}},\ }\href@noop {}
  {\  (\bibinfo {year} {2026})},\ \Eprint {https://arxiv.org/abs/2605.05290}
  {arXiv:2605.05290 [quant-ph]} \BibitemShut {NoStop}%
\bibitem [{\citenamefont {Bhattacharya}\ \emph {et~al.}(2022)\citenamefont
  {Bhattacharya}, \citenamefont {Nandy}, \citenamefont {Nath},\ and\
  \citenamefont {Sahu}}]{Bhattacharya:2022gbz}%
  \BibitemOpen
  \bibfield  {author} {\bibinfo {author} {\bibfnamefont {A.}~\bibnamefont
  {Bhattacharya}}, \bibinfo {author} {\bibfnamefont {P.}~\bibnamefont {Nandy}},
  \bibinfo {author} {\bibfnamefont {P.~P.}\ \bibnamefont {Nath}},\ and\
  \bibinfo {author} {\bibfnamefont {H.}~\bibnamefont {Sahu}},\ }\bibfield
  {title} {\bibinfo {title} {{Operator growth and Krylov construction in
  dissipative open quantum systems}},\ }\href
  {https://doi.org/10.1007/JHEP12(2022)081} {\bibfield  {journal} {\bibinfo
  {journal} {JHEP}\ }\textbf {\bibinfo {volume} {12}},\ \bibinfo {pages}
  {081}},\ \Eprint {https://arxiv.org/abs/2207.05347} {arXiv:2207.05347
  [quant-ph]} \BibitemShut {NoStop}%
\bibitem [{\citenamefont {Liu}\ \emph {et~al.}(2023)\citenamefont {Liu},
  \citenamefont {Tang},\ and\ \citenamefont {Zhai}}]{Liu:2022god}%
  \BibitemOpen
  \bibfield  {author} {\bibinfo {author} {\bibfnamefont {C.}~\bibnamefont
  {Liu}}, \bibinfo {author} {\bibfnamefont {H.}~\bibnamefont {Tang}},\ and\
  \bibinfo {author} {\bibfnamefont {H.}~\bibnamefont {Zhai}},\ }\bibfield
  {title} {\bibinfo {title} {{Krylov complexity in open quantum systems}},\
  }\href {https://doi.org/10.1103/PhysRevResearch.5.033085} {\bibfield
  {journal} {\bibinfo  {journal} {Phys. Rev. Res.}\ }\textbf {\bibinfo {volume}
  {5}},\ \bibinfo {pages} {033085} (\bibinfo {year} {2023})},\ \Eprint
  {https://arxiv.org/abs/2207.13603} {arXiv:2207.13603 [cond-mat.str-el]}
  \BibitemShut {NoStop}%
\bibitem [{\citenamefont {Bhattacharya}\ \emph {et~al.}(2023)\citenamefont
  {Bhattacharya}, \citenamefont {Nandy}, \citenamefont {Nath},\ and\
  \citenamefont {Sahu}}]{Bhattacharya:2023zqt}%
  \BibitemOpen
  \bibfield  {author} {\bibinfo {author} {\bibfnamefont {A.}~\bibnamefont
  {Bhattacharya}}, \bibinfo {author} {\bibfnamefont {P.}~\bibnamefont {Nandy}},
  \bibinfo {author} {\bibfnamefont {P.~P.}\ \bibnamefont {Nath}},\ and\
  \bibinfo {author} {\bibfnamefont {H.}~\bibnamefont {Sahu}},\ }\bibfield
  {title} {\bibinfo {title} {{On Krylov complexity in open systems: an approach
  via bi-Lanczos algorithm}},\ }\href {https://doi.org/10.1007/JHEP12(2023)066}
  {\bibfield  {journal} {\bibinfo  {journal} {JHEP}\ }\textbf {\bibinfo
  {volume} {12}},\ \bibinfo {pages} {066}},\ \Eprint
  {https://arxiv.org/abs/2303.04175} {arXiv:2303.04175 [quant-ph]} \BibitemShut
  {NoStop}%
\bibitem [{\citenamefont {Adhikari}\ \emph {et~al.}(2023)\citenamefont
  {Adhikari}, \citenamefont {Choudhury},\ and\ \citenamefont
  {Roy}}]{Adhikari:2022whf}%
  \BibitemOpen
  \bibfield  {author} {\bibinfo {author} {\bibfnamefont {K.}~\bibnamefont
  {Adhikari}}, \bibinfo {author} {\bibfnamefont {S.}~\bibnamefont
  {Choudhury}},\ and\ \bibinfo {author} {\bibfnamefont {A.}~\bibnamefont
  {Roy}},\ }\bibfield  {title} {\bibinfo {title} {{Krylov Complexity in Quantum
  Field Theory}},\ }\href {https://doi.org/10.1016/j.nuclphysb.2023.116263}
  {\bibfield  {journal} {\bibinfo  {journal} {Nucl. Phys. B}\ }\textbf
  {\bibinfo {volume} {993}},\ \bibinfo {pages} {116263} (\bibinfo {year}
  {2023})},\ \Eprint {https://arxiv.org/abs/2204.02250} {arXiv:2204.02250
  [hep-th]} \BibitemShut {NoStop}%
\bibitem [{\citenamefont {Avdoshkin}\ \emph {et~al.}(2024)\citenamefont
  {Avdoshkin}, \citenamefont {Dymarsky},\ and\ \citenamefont
  {Smolkin}}]{Avdoshkin:2022xuw}%
  \BibitemOpen
  \bibfield  {author} {\bibinfo {author} {\bibfnamefont {A.}~\bibnamefont
  {Avdoshkin}}, \bibinfo {author} {\bibfnamefont {A.}~\bibnamefont
  {Dymarsky}},\ and\ \bibinfo {author} {\bibfnamefont {M.}~\bibnamefont
  {Smolkin}},\ }\bibfield  {title} {\bibinfo {title} {{Krylov complexity in
  quantum field theory, and beyond}},\ }\href
  {https://doi.org/10.1007/JHEP06(2024)066} {\bibfield  {journal} {\bibinfo
  {journal} {JHEP}\ }\textbf {\bibinfo {volume} {06}},\ \bibinfo {pages}
  {066}},\ \Eprint {https://arxiv.org/abs/2212.14429} {arXiv:2212.14429
  [hep-th]} \BibitemShut {NoStop}%
\bibitem [{\citenamefont {Camargo}\ \emph {et~al.}(2023)\citenamefont
  {Camargo}, \citenamefont {Jahnke}, \citenamefont {Kim},\ and\ \citenamefont
  {Nishida}}]{Camargo:2022rnt}%
  \BibitemOpen
  \bibfield  {author} {\bibinfo {author} {\bibfnamefont {H.~A.}\ \bibnamefont
  {Camargo}}, \bibinfo {author} {\bibfnamefont {V.}~\bibnamefont {Jahnke}},
  \bibinfo {author} {\bibfnamefont {K.-Y.}\ \bibnamefont {Kim}},\ and\ \bibinfo
  {author} {\bibfnamefont {M.}~\bibnamefont {Nishida}},\ }\bibfield  {title}
  {\bibinfo {title} {{Krylov complexity in free and interacting scalar field
  theories with bounded power spectrum}},\ }\href
  {https://doi.org/10.1007/JHEP05(2023)226} {\bibfield  {journal} {\bibinfo
  {journal} {JHEP}\ }\textbf {\bibinfo {volume} {05}},\ \bibinfo {pages}
  {226}},\ \Eprint {https://arxiv.org/abs/2212.14702} {arXiv:2212.14702
  [hep-th]} \BibitemShut {NoStop}%
\bibitem [{\citenamefont {Vasli}\ \emph {et~al.}(2024)\citenamefont {Vasli},
  \citenamefont {Babaei~Velni}, \citenamefont {Mohammadi~Mozaffar},
  \citenamefont {Mollabashi},\ and\ \citenamefont
  {Alishahiha}}]{Vasli:2023syq}%
  \BibitemOpen
  \bibfield  {author} {\bibinfo {author} {\bibfnamefont {M.~J.}\ \bibnamefont
  {Vasli}}, \bibinfo {author} {\bibfnamefont {K.}~\bibnamefont {Babaei~Velni}},
  \bibinfo {author} {\bibfnamefont {M.~R.}\ \bibnamefont {Mohammadi~Mozaffar}},
  \bibinfo {author} {\bibfnamefont {A.}~\bibnamefont {Mollabashi}},\ and\
  \bibinfo {author} {\bibfnamefont {M.}~\bibnamefont {Alishahiha}},\ }\bibfield
   {title} {\bibinfo {title} {{Krylov complexity in Lifshitz-type scalar field
  theories}},\ }\href {https://doi.org/10.1140/epjc/s10052-024-12609-9}
  {\bibfield  {journal} {\bibinfo  {journal} {Eur. Phys. J. C}\ }\textbf
  {\bibinfo {volume} {84}},\ \bibinfo {pages} {235} (\bibinfo {year} {2024})},\
  \Eprint {https://arxiv.org/abs/2307.08307} {arXiv:2307.08307 [hep-th]}
  \BibitemShut {NoStop}%
\bibitem [{\citenamefont {He}\ and\ \citenamefont {Zhang}(2024)}]{He:2024xjp}%
  \BibitemOpen
  \bibfield  {author} {\bibinfo {author} {\bibfnamefont {P.-Z.}\ \bibnamefont
  {He}}\ and\ \bibinfo {author} {\bibfnamefont {H.-Q.}\ \bibnamefont {Zhang}},\
  }\bibfield  {title} {\bibinfo {title} {{Probing Krylov complexity in scalar
  field theory with general temperatures}},\ }\href
  {https://doi.org/10.1007/JHEP11(2024)014} {\bibfield  {journal} {\bibinfo
  {journal} {JHEP}\ }\textbf {\bibinfo {volume} {11}},\ \bibinfo {pages}
  {014}},\ \Eprint {https://arxiv.org/abs/2407.02756} {arXiv:2407.02756
  [hep-th]} \BibitemShut {NoStop}%
\bibitem [{\citenamefont {He}\ and\ \citenamefont {Zhang}(2025)}]{He:2024hkw}%
  \BibitemOpen
  \bibfield  {author} {\bibinfo {author} {\bibfnamefont {P.-Z.}\ \bibnamefont
  {He}}\ and\ \bibinfo {author} {\bibfnamefont {H.-Q.}\ \bibnamefont {Zhang}},\
  }\bibfield  {title} {\bibinfo {title} {{Krylov complexity in the
  Schr{\"o}dinger field theory}},\ }\href
  {https://doi.org/10.1007/JHEP03(2025)142} {\bibfield  {journal} {\bibinfo
  {journal} {JHEP}\ }\textbf {\bibinfo {volume} {03}},\ \bibinfo {pages}
  {142}},\ \Eprint {https://arxiv.org/abs/2411.16302} {arXiv:2411.16302
  [hep-th]} \BibitemShut {NoStop}%
\bibitem [{\citenamefont {He}\ \emph {et~al.}(2026)\citenamefont {He},
  \citenamefont {Liu}, \citenamefont {Zhang},\ and\ \citenamefont
  {Jiang}}]{He:2025guu}%
  \BibitemOpen
  \bibfield  {author} {\bibinfo {author} {\bibfnamefont {P.-Z.}\ \bibnamefont
  {He}}, \bibinfo {author} {\bibfnamefont {L.-H.}\ \bibnamefont {Liu}},
  \bibinfo {author} {\bibfnamefont {H.-Q.}\ \bibnamefont {Zhang}},\ and\
  \bibinfo {author} {\bibfnamefont {Q.-Q.}\ \bibnamefont {Jiang}},\ }\bibfield
  {title} {\bibinfo {title} {{Krylov complexity and Wightman power spectrum
  with positive chemical potential in Schr{\"o}dinger field theory}},\ }\href
  {https://doi.org/10.1007/JHEP02(2026)259} {\bibfield  {journal} {\bibinfo
  {journal} {JHEP}\ }\textbf {\bibinfo {volume} {02}},\ \bibinfo {pages}
  {259}},\ \Eprint {https://arxiv.org/abs/2509.14742} {arXiv:2509.14742
  [hep-th]} \BibitemShut {NoStop}%
\bibitem [{\citenamefont {Aguilar-Gutierrez}\ and\ \citenamefont
  {Rolph}(2024)}]{Aguilar-Gutierrez:2023nyk}%
  \BibitemOpen
  \bibfield  {author} {\bibinfo {author} {\bibfnamefont {S.~E.}\ \bibnamefont
  {Aguilar-Gutierrez}}\ and\ \bibinfo {author} {\bibfnamefont {A.}~\bibnamefont
  {Rolph}},\ }\bibfield  {title} {\bibinfo {title} {{Krylov complexity is not a
  measure of distance between states or operators}},\ }\href
  {https://doi.org/10.1103/PhysRevD.109.L081701} {\bibfield  {journal}
  {\bibinfo  {journal} {Phys. Rev. D}\ }\textbf {\bibinfo {volume} {109}},\
  \bibinfo {pages} {L081701} (\bibinfo {year} {2024})},\ \Eprint
  {https://arxiv.org/abs/2311.04093} {arXiv:2311.04093 [hep-th]} \BibitemShut
  {NoStop}%
\bibitem [{\citenamefont {Caputa}\ \emph {et~al.}(2024)\citenamefont {Caputa},
  \citenamefont {Jeong}, \citenamefont {Liu}, \citenamefont {Pedraza},\ and\
  \citenamefont {Qu}}]{Caputa:2024vrn}%
  \BibitemOpen
  \bibfield  {author} {\bibinfo {author} {\bibfnamefont {P.}~\bibnamefont
  {Caputa}}, \bibinfo {author} {\bibfnamefont {H.-S.}\ \bibnamefont {Jeong}},
  \bibinfo {author} {\bibfnamefont {S.}~\bibnamefont {Liu}}, \bibinfo {author}
  {\bibfnamefont {J.~F.}\ \bibnamefont {Pedraza}},\ and\ \bibinfo {author}
  {\bibfnamefont {L.-C.}\ \bibnamefont {Qu}},\ }\bibfield  {title} {\bibinfo
  {title} {{Krylov complexity of density matrix operators}},\ }\href
  {https://doi.org/10.1007/JHEP05(2024)337} {\bibfield  {journal} {\bibinfo
  {journal} {JHEP}\ }\textbf {\bibinfo {volume} {05}},\ \bibinfo {pages}
  {337}},\ \Eprint {https://arxiv.org/abs/2402.09522} {arXiv:2402.09522
  [hep-th]} \BibitemShut {NoStop}%
\bibitem [{\citenamefont {Seetharaman}\ \emph {et~al.}(2025)\citenamefont
  {Seetharaman}, \citenamefont {Singh},\ and\ \citenamefont
  {Nath}}]{Seetharaman:2024ket}%
  \BibitemOpen
  \bibfield  {author} {\bibinfo {author} {\bibfnamefont {S.}~\bibnamefont
  {Seetharaman}}, \bibinfo {author} {\bibfnamefont {C.}~\bibnamefont {Singh}},\
  and\ \bibinfo {author} {\bibfnamefont {R.}~\bibnamefont {Nath}},\ }\bibfield
  {title} {\bibinfo {title} {{Properties of Krylov state complexity in qubit
  dynamics}},\ }\href {https://doi.org/10.1103/PhysRevD.111.076014} {\bibfield
  {journal} {\bibinfo  {journal} {Phys. Rev. D}\ }\textbf {\bibinfo {volume}
  {111}},\ \bibinfo {pages} {076014} (\bibinfo {year} {2025})},\ \Eprint
  {https://arxiv.org/abs/2407.21776} {arXiv:2407.21776 [quant-ph]} \BibitemShut
  {NoStop}%
\bibitem [{\citenamefont {Li}\ and\ \citenamefont
  {Liu}(2024{\natexlab{a}})}]{Li:2024kfm}%
  \BibitemOpen
  \bibfield  {author} {\bibinfo {author} {\bibfnamefont {T.}~\bibnamefont
  {Li}}\ and\ \bibinfo {author} {\bibfnamefont {L.-H.}\ \bibnamefont {Liu}},\
  }\bibfield  {title} {\bibinfo {title} {{Inflationary Krylov complexity}},\
  }\href {https://doi.org/10.1007/JHEP04(2024)123} {\bibfield  {journal}
  {\bibinfo  {journal} {JHEP}\ }\textbf {\bibinfo {volume} {04}},\ \bibinfo
  {pages} {123}},\ \Eprint {https://arxiv.org/abs/2401.09307} {arXiv:2401.09307
  [hep-th]} \BibitemShut {NoStop}%
\bibitem [{\citenamefont {Li}\ and\ \citenamefont
  {Liu}(2024{\natexlab{b}})}]{Li:2024iji}%
  \BibitemOpen
  \bibfield  {author} {\bibinfo {author} {\bibfnamefont {T.}~\bibnamefont
  {Li}}\ and\ \bibinfo {author} {\bibfnamefont {L.-H.}\ \bibnamefont {Liu}},\
  }\bibfield  {title} {\bibinfo {title} {{Inflationary complexity of thermal
  state}},\ }\href@noop {} {\  (\bibinfo {year} {2024}{\natexlab{b}})},\
  \Eprint {https://arxiv.org/abs/2405.01433} {arXiv:2405.01433 [hep-th]}
  \BibitemShut {NoStop}%
\bibitem [{\citenamefont {Zhai}\ and\ \citenamefont
  {Liu}(2026)}]{Zhai:2024odw}%
  \BibitemOpen
  \bibfield  {author} {\bibinfo {author} {\bibfnamefont {K.-H.}\ \bibnamefont
  {Zhai}}\ and\ \bibinfo {author} {\bibfnamefont {L.-H.}\ \bibnamefont {Liu}},\
  }\bibfield  {title} {\bibinfo {title} {{Krylov Complexity in Early
  Universe}},\ }\href {https://doi.org/10.1093/ptep/ptag012} {\bibfield
  {journal} {\bibinfo  {journal} {PTEP}\ }\textbf {\bibinfo {volume} {2026}},\
  \bibinfo {pages} {023E04} (\bibinfo {year} {2026})},\ \Eprint
  {https://arxiv.org/abs/2411.18405} {arXiv:2411.18405 [hep-th]} \BibitemShut
  {NoStop}%
\bibitem [{\citenamefont {Craps}\ \emph {et~al.}(2024)\citenamefont {Craps},
  \citenamefont {Evnin},\ and\ \citenamefont {Pascuzzi}}]{Craps:2023ivc}%
  \BibitemOpen
  \bibfield  {author} {\bibinfo {author} {\bibfnamefont {B.}~\bibnamefont
  {Craps}}, \bibinfo {author} {\bibfnamefont {O.}~\bibnamefont {Evnin}},\ and\
  \bibinfo {author} {\bibfnamefont {G.}~\bibnamefont {Pascuzzi}},\ }\bibfield
  {title} {\bibinfo {title} {{A Relation between Krylov and Nielsen
  Complexity}},\ }\href {https://doi.org/10.1103/PhysRevLett.132.160402}
  {\bibfield  {journal} {\bibinfo  {journal} {Phys. Rev. Lett.}\ }\textbf
  {\bibinfo {volume} {132}},\ \bibinfo {pages} {160402} (\bibinfo {year}
  {2024})},\ \Eprint {https://arxiv.org/abs/2311.18401} {arXiv:2311.18401
  [quant-ph]} \BibitemShut {NoStop}%
\bibitem [{\citenamefont {Craps}\ \emph {et~al.}(2025)\citenamefont {Craps},
  \citenamefont {Pascuzzi}, \citenamefont {Pedraza}, \citenamefont {Qu},\ and\
  \citenamefont {Ruan}}]{Craps:2025kub}%
  \BibitemOpen
  \bibfield  {author} {\bibinfo {author} {\bibfnamefont {B.}~\bibnamefont
  {Craps}}, \bibinfo {author} {\bibfnamefont {G.}~\bibnamefont {Pascuzzi}},
  \bibinfo {author} {\bibfnamefont {J.~F.}\ \bibnamefont {Pedraza}}, \bibinfo
  {author} {\bibfnamefont {L.-C.}\ \bibnamefont {Qu}},\ and\ \bibinfo {author}
  {\bibfnamefont {S.-M.}\ \bibnamefont {Ruan}},\ }\bibfield  {title} {\bibinfo
  {title} {{Explicit Connections Between Krylov and Nielsen Complexity}},\
  }\href@noop {} {\  (\bibinfo {year} {2025})},\ \Eprint
  {https://arxiv.org/abs/2511.15799} {arXiv:2511.15799 [hep-th]} \BibitemShut
  {NoStop}%
\bibitem [{\citenamefont {Lv}\ \emph {et~al.}(2024)\citenamefont {Lv},
  \citenamefont {Zhang},\ and\ \citenamefont {Zhou}}]{Lv:2023jbv}%
  \BibitemOpen
  \bibfield  {author} {\bibinfo {author} {\bibfnamefont {C.}~\bibnamefont
  {Lv}}, \bibinfo {author} {\bibfnamefont {R.}~\bibnamefont {Zhang}},\ and\
  \bibinfo {author} {\bibfnamefont {Q.}~\bibnamefont {Zhou}},\ }\bibfield
  {title} {\bibinfo {title} {{Building Krylov complexity from circuit
  complexity}},\ }\href {https://doi.org/10.1103/PhysRevResearch.6.L042001}
  {\bibfield  {journal} {\bibinfo  {journal} {Phys. Rev. Res.}\ }\textbf
  {\bibinfo {volume} {6}},\ \bibinfo {pages} {L042001} (\bibinfo {year}
  {2024})},\ \Eprint {https://arxiv.org/abs/2303.07343} {arXiv:2303.07343
  [quant-ph]} \BibitemShut {NoStop}%
\bibitem [{\citenamefont {Viswanath}\ and\ \citenamefont
  {M{\"u}ller}(1994)}]{viswanath1994recursion}%
  \BibitemOpen
  \bibfield  {author} {\bibinfo {author} {\bibfnamefont {V.}~\bibnamefont
  {Viswanath}}\ and\ \bibinfo {author} {\bibfnamefont {G.}~\bibnamefont
  {M{\"u}ller}},\ }\href@noop {} {\emph {\bibinfo {title} {The recursion
  method:application to many-body dynamics}}}\ (\bibinfo  {publisher}
  {Springer},\ \bibinfo {year} {1994})\BibitemShut {NoStop}%
\bibitem [{\citenamefont {Takahashi}\ and\ \citenamefont {del
  Campo}(2025)}]{Takahashi:2024hex}%
  \BibitemOpen
  \bibfield  {author} {\bibinfo {author} {\bibfnamefont {K.}~\bibnamefont
  {Takahashi}}\ and\ \bibinfo {author} {\bibfnamefont {A.}~\bibnamefont {del
  Campo}},\ }\bibfield  {title} {\bibinfo {title} {{Krylov Subspace Methods for
  Quantum Dynamics with Time-Dependent Generators}},\ }\href
  {https://doi.org/10.1103/PhysRevLett.134.030401} {\bibfield  {journal}
  {\bibinfo  {journal} {Phys. Rev. Lett.}\ }\textbf {\bibinfo {volume} {134}},\
  \bibinfo {pages} {030401} (\bibinfo {year} {2025})},\ \Eprint
  {https://arxiv.org/abs/2408.08383} {arXiv:2408.08383 [quant-ph]} \BibitemShut
  {NoStop}%
\bibitem [{\citenamefont {Howland}(1974)}]{Howland:1974aa}%
  \BibitemOpen
  \bibfield  {author} {\bibinfo {author} {\bibfnamefont {J.~S.}\ \bibnamefont
  {Howland}},\ }\bibfield  {title} {\bibinfo {title} {Stationary scattering
  theory for time-dependent hamiltonians},\ }\href
  {https://doi.org/10.1007/BF01351346} {\bibfield  {journal} {\bibinfo
  {journal} {Mathematische Annalen}\ }\textbf {\bibinfo {volume} {207}},\
  \bibinfo {pages} {315} (\bibinfo {year} {1974})}\BibitemShut {NoStop}%
\bibitem [{\citenamefont {Peskin}\ and\ \citenamefont
  {Moiseyev}(1993)}]{Peskin:1993}%
  \BibitemOpen
  \bibfield  {author} {\bibinfo {author} {\bibfnamefont {U.}~\bibnamefont
  {Peskin}}\ and\ \bibinfo {author} {\bibfnamefont {N.}~\bibnamefont
  {Moiseyev}},\ }\bibfield  {title} {\bibinfo {title} {The solution of the
  time‐dependent schrodinger equation by the (t,t') method: Theory,
  computational algorithm and applications},\ }\href
  {https://doi.org/10.1063/1.466058} {\bibfield  {journal} {\bibinfo  {journal}
  {The Journal of Chemical Physics}\ }\textbf {\bibinfo {volume} {99}},\
  \bibinfo {pages} {4590} (\bibinfo {year} {1993})}\BibitemShut {NoStop}%
\bibitem [{\citenamefont {Peskin}\ \emph {et~al.}(1994)\citenamefont {Peskin},
  \citenamefont {Kosloff},\ and\ \citenamefont {Moiseyev}}]{Peskin:1994}%
  \BibitemOpen
  \bibfield  {author} {\bibinfo {author} {\bibfnamefont {U.}~\bibnamefont
  {Peskin}}, \bibinfo {author} {\bibfnamefont {R.}~\bibnamefont {Kosloff}},\
  and\ \bibinfo {author} {\bibfnamefont {N.}~\bibnamefont {Moiseyev}},\
  }\bibfield  {title} {\bibinfo {title} {The solution of the time dependent
  schr{\"o}dinger equation by the (t,t') method: The use of global polynomial
  propagators for time dependent hamiltonians},\ }\href
  {https://doi.org/10.1063/1.466739} {\bibfield  {journal} {\bibinfo  {journal}
  {The Journal of Chemical Physics}\ }\textbf {\bibinfo {volume} {100}},\
  \bibinfo {pages} {8849} (\bibinfo {year} {1994})}\BibitemShut {NoStop}%
\bibitem [{\citenamefont {Faraji~Astaneh}\ and\ \citenamefont
  {Kafashi}(2026)}]{FarajiAstaneh:2026aks}%
  \BibitemOpen
  \bibfield  {author} {\bibinfo {author} {\bibfnamefont {A.}~\bibnamefont
  {Faraji~Astaneh}}\ and\ \bibinfo {author} {\bibfnamefont {P.}~\bibnamefont
  {Kafashi}},\ }\bibfield  {title} {\bibinfo {title} {{Krylov Complexity for
  Time-Dependent Hamiltonians}},\ }\href@noop {} {\  (\bibinfo {year}
  {2026})},\ \Eprint {https://arxiv.org/abs/2607.10454} {arXiv:2607.10454
  [hep-th]} \BibitemShut {NoStop}%
\bibitem [{\citenamefont {Berry}(1984)}]{Berry:1984jv}%
  \BibitemOpen
  \bibfield  {author} {\bibinfo {author} {\bibfnamefont {M.~V.}\ \bibnamefont
  {Berry}},\ }\bibfield  {title} {\bibinfo {title} {{Quantal phase factors
  accompanying adiabatic changes}},\ }\href
  {https://doi.org/10.1098/rspa.1984.0023} {\bibfield  {journal} {\bibinfo
  {journal} {Proc. Roy. Soc. Lond. A}\ }\textbf {\bibinfo {volume} {392}},\
  \bibinfo {pages} {45} (\bibinfo {year} {1984})}\BibitemShut {NoStop}%
\bibitem [{\citenamefont {Shapere}\ and\ \citenamefont
  {Wilczek}(1989)}]{shapere1989geometric}%
  \BibitemOpen
  \bibfield  {author} {\bibinfo {author} {\bibfnamefont {A.}~\bibnamefont
  {Shapere}}\ and\ \bibinfo {author} {\bibfnamefont {F.}~\bibnamefont
  {Wilczek}},\ }\href@noop {} {\emph {\bibinfo {title} {Geometric phases in
  physics}}},\ Vol.~\bibinfo {volume} {5}\ (\bibinfo  {publisher} {World
  scientific},\ \bibinfo {year} {1989})\BibitemShut {NoStop}%
\bibitem [{\citenamefont {Thouless}\ \emph {et~al.}(1982)\citenamefont
  {Thouless}, \citenamefont {Kohmoto}, \citenamefont {Nightingale},\ and\
  \citenamefont {den Nijs}}]{Thouless:1982zz}%
  \BibitemOpen
  \bibfield  {author} {\bibinfo {author} {\bibfnamefont {D.~J.}\ \bibnamefont
  {Thouless}}, \bibinfo {author} {\bibfnamefont {M.}~\bibnamefont {Kohmoto}},
  \bibinfo {author} {\bibfnamefont {M.~P.}\ \bibnamefont {Nightingale}},\ and\
  \bibinfo {author} {\bibfnamefont {M.}~\bibnamefont {den Nijs}},\ }\bibfield
  {title} {\bibinfo {title} {{Quantized Hall Conductance in a Two-Dimensional
  Periodic Potential}},\ }\href {https://doi.org/10.1103/PhysRevLett.49.405}
  {\bibfield  {journal} {\bibinfo  {journal} {Phys. Rev. Lett.}\ }\textbf
  {\bibinfo {volume} {49}},\ \bibinfo {pages} {405} (\bibinfo {year}
  {1982})}\BibitemShut {NoStop}%
\bibitem [{\citenamefont {Kohmoto}(1985)}]{KOHMOTO1985343}%
  \BibitemOpen
  \bibfield  {author} {\bibinfo {author} {\bibfnamefont {M.}~\bibnamefont
  {Kohmoto}},\ }\bibfield  {title} {\bibinfo {title} {Topological invariant and
  the quantization of the hall conductance},\ }\href
  {https://doi.org/https://doi.org/10.1016/0003-4916(85)90148-4} {\bibfield
  {journal} {\bibinfo  {journal} {Annals of Physics}\ }\textbf {\bibinfo
  {volume} {160}},\ \bibinfo {pages} {343} (\bibinfo {year}
  {1985})}\BibitemShut {NoStop}%
\bibitem [{\citenamefont {Jungwirth}\ \emph {et~al.}(2002)\citenamefont
  {Jungwirth}, \citenamefont {Niu},\ and\ \citenamefont
  {MacDonald}}]{Jungwirth:2002zz}%
  \BibitemOpen
  \bibfield  {author} {\bibinfo {author} {\bibfnamefont {T.}~\bibnamefont
  {Jungwirth}}, \bibinfo {author} {\bibfnamefont {Q.}~\bibnamefont {Niu}},\
  and\ \bibinfo {author} {\bibfnamefont {A.~H.}\ \bibnamefont {MacDonald}},\
  }\bibfield  {title} {\bibinfo {title} {{Anomalous Hall Effect in
  Ferromagnetic Semiconductors}},\ }\href
  {https://doi.org/10.1103/PhysRevLett.88.207208} {\bibfield  {journal}
  {\bibinfo  {journal} {Phys. Rev. Lett.}\ }\textbf {\bibinfo {volume} {88}},\
  \bibinfo {pages} {207208} (\bibinfo {year} {2002})},\ \Eprint
  {https://arxiv.org/abs/cond-mat/0110484} {arXiv:cond-mat/0110484}
  \BibitemShut {NoStop}%
\bibitem [{\citenamefont {Hasan}\ and\ \citenamefont
  {Kane}(2010)}]{RevModPhys.82.3045}%
  \BibitemOpen
  \bibfield  {author} {\bibinfo {author} {\bibfnamefont {M.~Z.}\ \bibnamefont
  {Hasan}}\ and\ \bibinfo {author} {\bibfnamefont {C.~L.}\ \bibnamefont
  {Kane}},\ }\bibfield  {title} {\bibinfo {title} {Colloquium: Topological
  insulators},\ }\href {https://doi.org/10.1103/RevModPhys.82.3045} {\bibfield
  {journal} {\bibinfo  {journal} {Rev. Mod. Phys.}\ }\textbf {\bibinfo {volume}
  {82}},\ \bibinfo {pages} {3045} (\bibinfo {year} {2010})}\BibitemShut
  {NoStop}%
\bibitem [{\citenamefont {Xiao}\ \emph {et~al.}(2010)\citenamefont {Xiao},
  \citenamefont {Chang},\ and\ \citenamefont {Niu}}]{Xiao:2009rm}%
  \BibitemOpen
  \bibfield  {author} {\bibinfo {author} {\bibfnamefont {D.}~\bibnamefont
  {Xiao}}, \bibinfo {author} {\bibfnamefont {M.-C.}\ \bibnamefont {Chang}},\
  and\ \bibinfo {author} {\bibfnamefont {Q.}~\bibnamefont {Niu}},\ }\bibfield
  {title} {\bibinfo {title} {{Berry Phase Effects on Electronic Properties}},\
  }\href {https://doi.org/10.1103/RevModPhys.82.1959} {\bibfield  {journal}
  {\bibinfo  {journal} {Rev. Mod. Phys.}\ }\textbf {\bibinfo {volume} {82}},\
  \bibinfo {pages} {1959} (\bibinfo {year} {2010})},\ \Eprint
  {https://arxiv.org/abs/0907.2021} {arXiv:0907.2021 [cond-mat.mes-hall]}
  \BibitemShut {NoStop}%
\bibitem [{\citenamefont {Sakurai}\ and\ \citenamefont
  {Napolitano}(2020)}]{Sakurai:2011zz}%
  \BibitemOpen
  \bibfield  {author} {\bibinfo {author} {\bibfnamefont {J.~J.}\ \bibnamefont
  {Sakurai}}\ and\ \bibinfo {author} {\bibfnamefont {J.}~\bibnamefont
  {Napolitano}},\ }\href {https://doi.org/10.1017/9781108587280} {\emph
  {\bibinfo {title} {{Modern Quantum Mechanics}}}},\ \bibinfo {edition} {3rd}\
  ed.,\ Quantum physics, quantum information and quantum computation\ (\bibinfo
   {publisher} {Cambridge University Press},\ \bibinfo {year}
  {2020})\BibitemShut {NoStop}%
\bibitem [{SM()}]{SM}%
  \BibitemOpen
  \href@noop {} {\bibinfo {title} {Please refer to the suppelemental materials
  for details.}}\BibitemShut {Stop}%
\bibitem [{Note1()}]{Note1}%
  \BibitemOpen
  \bibinfo {note} {For the azimuthal angle $\varphi (t)$, we always set it
  time-dependent. What we mean the constant parameter for $\varphi (t)$ is that
  its angular velocity $\omega =\protect \dot \varphi (t)$ is a constant rather
  than $\varphi (t)$ itself.}\BibitemShut {Stop}%
\bibitem [{\citenamefont {Rabi}(1937)}]{Rabi:1937dgo}%
  \BibitemOpen
  \bibfield  {author} {\bibinfo {author} {\bibfnamefont {I.~I.}\ \bibnamefont
  {Rabi}},\ }\bibfield  {title} {\bibinfo {title} {{Space Quantization in a
  Gyrating Magnetic Field}},\ }\href {https://doi.org/10.1103/PhysRev.51.652}
  {\bibfield  {journal} {\bibinfo  {journal} {Phys. Rev.}\ }\textbf {\bibinfo
  {volume} {51}},\ \bibinfo {pages} {652} (\bibinfo {year} {1937})}\BibitemShut
  {NoStop}%
\bibitem [{Note2()}]{Note2}%
  \BibitemOpen
  \bibinfo {note} {We should stress that ${\protect \bf A}(t)$ is not the Berry
  connection matrix from the instantaneous eigenstates of original Hamiltonian
  \protect \textup {\hbox {\mathsurround \z@ \protect \normalfont
  (\ignorespaces \ref {hamiltonian}\unskip \@@italiccorr )}}, but rather it is
  from the instantaneous eigenstates of the effective Hamiltonian $H_{\protect
  \text {eff}}$ composed of the Lanczos coefficients.}\BibitemShut {Stop}%
\bibitem [{\citenamefont {Aharonov}\ and\ \citenamefont
  {Bohm}(1959)}]{Aharonov:1959fk}%
  \BibitemOpen
  \bibfield  {author} {\bibinfo {author} {\bibfnamefont {Y.}~\bibnamefont
  {Aharonov}}\ and\ \bibinfo {author} {\bibfnamefont {D.}~\bibnamefont
  {Bohm}},\ }\bibfield  {title} {\bibinfo {title} {{Significance of
  electromagnetic potentials in the quantum theory}},\ }\href
  {https://doi.org/10.1103/PhysRev.115.485} {\bibfield  {journal} {\bibinfo
  {journal} {Phys. Rev.}\ }\textbf {\bibinfo {volume} {115}},\ \bibinfo {pages}
  {485} (\bibinfo {year} {1959})}\BibitemShut {NoStop}%
\end{thebibliography}%

\end{document}